\documentclass[a4paper,12pt]{article}
\pdfoutput=1
\usepackage{epsfig,graphicx,xcolor,soul,amsbsy,amssymb,latexsym,amsfonts,amsmath,tcolorbox,setspace}
\usepackage{pstricks}
\usepackage{color}
\usepackage[numbers,square,comma, compress]{natbib}

\usepackage{eurosym}
\usepackage[a4paper]{geometry}
\usepackage{graphicx}
\usepackage{tikz}
\usepackage{xcolor}
\usepackage{amsmath,amssymb,amsfonts,tcolorbox,pstricks,setspace}

\numberwithin{equation}{section}

\newcommand{\bea}{\begin{eqnarray}\displaystyle}
\newcommand{\eea}{\end{eqnarray}}

\newcommand{\figref}[1]{Fig.~\protect\ref{#1}}

\title{
\vspace{-1.8cm}
\bf{Self-Duality and Self-Similarity \\
of \\
Little String Orbifolds}\\[15pt]}
\author{\large \textsc{Stefan~Hohenegger\footnote{\tt s.hohenegger@ipnl.in2p3.fr}~,~Amer Iqbal\footnote{\tt  amer@alum.mit.edu}~,~ Soo-Jong Rey\,\footnote{\tt sjrey@snu.ac.kr}}}
\date{}

\begin{document}

\maketitle

\begin{center}
\renewcommand{\thefootnote}{\fnsymbol{footnote}}\vspace{-0.5cm}
${}^{\footnotemark[1]}$ Universit\'e de Lyon\\
UMR 5822, CNRS/IN2P3, Institut de Physique Nucl\'eaire de Lyon\\ 4 rue Enrico Fermi, 69622 Villeurbanne Cedex, \rm FRANCE\\[0.4cm]
${}^{\footnotemark[2]}$ Abdus Salam School of Mathematical Sciences \\ Government College University, Lahore, PAKISTAN\\[0.4cm]
${}^{\footnotemark[2]}$ Center for Theoretical Physics, Lahore, PAKISTAN\\[0.4cm]
${}^{\footnotemark[3]}$ School of Physics and Astronomy \& Center for Theoretical Physics\\
          Seoul National University, Seoul 08826 \rm KOREA\\[0.4cm]
${}^{\footnotemark[3]}$ B.W. Lee Center for Fields, Gravity \& Strings \\
Institute for Basic Sciences, Daejeon 34047 \rm KOREA\\[1cm]
\end{center}

\begin{abstract}
\noindent
We study a class of ${\cal N}=(1,0)$ little string theories obtained from orbifolds of M-brane configurations. These are realised in two different ways that are dual to each other: either as $M$ parallel M5-branes probing a transverse $A_{N-1}$ singularity or $N$ M5-branes probing an $A_{M-1}$ singularity. These backgrounds can further be dualised into toric, non-compact Calabi-Yau threefolds $X_{N,M}$ which have double elliptic fibrations and thus give a natural geometric description of T-duality of the little string theories. The little string partition functions are captured by the topological string partition function of $X_{N,M}$. We analyse in detail the free energies $\Sigma_{N,M}$ associated with the latter in a special region in the K\"ahler moduli space of $X_{N,M}$ and discover a remarkable property: in the Nekrasov-Shatashvili-limit, $\Sigma_{N,M}$ is identical to $NM$ times $\Sigma_{1,1}$. This entails that the BPS degeneracies for any $(N,M)$ can uniquely be reconstructed from the $(N,M)=(1,1)$ configuration,  a property we refer to as self-similarity. Moreover, as $\Sigma_{1,1}$ is known to display a number of recursive structures, BPS degeneracies of little string configurations for arbitrary $(N,M)$ as well acquire additional symmetries. These symmetries suggest that in this special region the two little string theories described above are self-dual under T-duality.

\end{abstract}

\tableofcontents

\onehalfspacing
\vskip1cm

\newpage

\section{Introduction}
The recent study of BPS states in certain configurations of M-branes turned out to be very fruitful and beneficial from a variety of perspectives. On the one hand, the discovery of very rich and interesting structures shed new light on the inner workings of M-theory and its basic building blocks. On the other hand, with the help of dualities, these new findings also unravelled new structures and symmetries in related physical theories, for example supersymmetric gauge theories and, most recently, little string theories from various viewpoints
\cite{Witten:1995zh, Aspinwall:1997ye, Seiberg:1997zk, Intriligator:1997dh, Hanany:1997gh, Brunner:1997gf}  (see \cite{Aharony:1999ks,Kutasov:2001uf} for reviews).

More concretely, in \cite{Haghighat:2013gba,Haghighat:2013tka,Hohenegger:2013ala,Hohenegger:2015cba}, a stack of $N$ 
parallel M5-branes with M2-branes stretched between them was studied. In \cite{Hohenegger:2015btj}, this setup was generalised to the point that a transverse as well as a parallel direction to the M5-branes are compactified on $\mathbb{S}^1_{\tau}\times \mathbb{S}^1_{\rho}$ (with radii $\tau$ and $\rho$ , respectively). The partition functions $\mathcal{Z}_{X_{N,1}}$ of such setups that count BPS states can be obtained efficiently through a dual description in terms of non-compact toric Calabi-Yau manifolds $X_{N,1}$, with the help of the (refined) topological vertex (see \cite{Aganagic:2003db,Hollowood:2003cv,Iqbal:2007ii}). The compactification of the M-brane setup is reflected in a double fibration structure of $X_{N,1}$ with elliptic parameters ($\rho,\tau)$, respectively.  As discussed in \cite{Hohenegger:2015btj}, $\mathcal{Z}_{X_{N,1}}$ is related to the partition functions of two different (but dual) supersymmetric gauge theories that can be associated to the toric Calabi-Yau threefold $X_{N,1}$. Here $\mathcal{Z}_{X_{N,1}}$ is regularised by formulating the theories on the $\Omega$-background, parameterised by $\epsilon_{1,2}\in\mathbb{R}$. Furthermore, these two theories in turn can be understood as gauge theory descriptions of $\mathbb{S}^1$-compactifications of $\mathcal{N}=(1,1) $ type IIa and $\mathcal{N}=(2,0)$ type IIb little string theory, respectively.

The high degree of symmetry of the M-brane configuration (and thus also of $X_{N,1}$) is also reflected in the corresponding partition functions. Notably, it was conjectured in \cite{Bhardwaj:2015oru} that the exchange of the two elliptic parameters $\rho\longleftrightarrow\tau$ in these configurations, corresponds to
T-duality from the point of view of the little string theories. This was shown in \cite{Hohenegger:2015btj} by relating the topological string partition functions of double elliptically fibered Calabi-Yau threefolds to the partition functions of T-dual pairs of little string theories. In \cite{Bhardwaj:2015oru}, the connection between T-duality of little strings and double elliptic fibrations was discussed in generality by classifying little string theories via geometric phases in F-theory. In the cases discussed above, upon denoting the type IIa and IIb partition functions $Z_{\text{IIa}}$ and $Z_{\text{IIb}}$ respectively, the connection to $X_{N,1}$ suggests~\cite{Kim:2015gha, Bhardwaj:2015oru, Hohenegger:2015btj}
\begin{align}
Z_{\text{IIa}}(\tau,\rho)=\mathcal{Z}_{X_{N,1}}(\tau, \rho)=Z_{\text{IIb}}(\rho,\tau)\,,
\end{align}
thus making T-duality manifest. This conjecture is supported by the fact that the counting of certain BPS states in the two dual theories (extracted from $\mathcal{Z}_{X_{N,1}}$ in the form of the free energy), indeed agrees \cite{Hohenegger:2015btj} for specific M2-brane configurations.

In this paper, we extend the analysis of these dualities to a class of little string theories with $\mathcal{N}$=(1,0) supersymmetry, which are obtained through an orbifold of the original brane configuration that  turns the transverse space into an $\text{ALE}_{A_{M-1}}$ space. These brane configurations can again be dualised to a toric non-compact Calabi-Yau threefold denoted $X_{N,M}$. This possibility was first discussed in \cite{Haghighat:2013tka,Hohenegger:2013ala} and the corresponding partition functions $\mathcal{Z}_{X_{N,M}}$ computed using the (refined) topological vertex formalism. These configurations depend on two sets of K\"ahler parameters $T_{i=1,\ldots,M}$ and $t_{a=1\ldots,N}$. Physically, they correspond to the distances between the (stacks of) branes, since due to the orbifold it is now also possible to separate the M2-branes (see \cite{Haghighat:2013tka}). Due to the compactification of the system on $\mathbb{S}^1_{\tau}\times \mathbb{S}^1_{\rho}$ and decoupling of gravity, these dynamical parameters are not all independent, but rather sum up to $\tau$ and $\rho$ respectively.

Motivated by the viewpoint of a sigma model description for the elliptic genus (see \cite{EG} for a definition), we observe a number of remarkable properties of the partition function $\mathcal{Z}_{X_{N,M}}$ as well as the associated free energy, $\Sigma_{N, M} = \mbox{Plog} \, \mathcal{Z}_{X_{N, M}}$, when
\begin{align}
&T_1=T_2=\ldots=T_M=-\frac{2\pi i}{M}\,\tau &&\text{and}&&t_1=t_2=\ldots=t_N=-\frac{2\pi i}{N}\,\rho \,,\label{PointSimGenIntro}
\end{align}
where all neighbouring branes are separated by an equal distance (notice, however, that we still keep the finite parameters $\tau\neq \rho$ in general). At (\ref{PointSimGenIntro}) and in the Nekrasov-Shatashvili (NS) limit $\epsilon_2\to 0$ (see \cite{Nekrasov:2009rc, Mironov:2009uv}), we observe\footnote{We obtain this result by comparing the first terms in a Fourier expansion with respect to the remaining physical parameters $Q_\tau=e^{2\pi i\tau}$ and $Q_\rho=e^{2\pi i\rho}$ and make a conjecture based on the emergent combinatoric patterns.} that the free energies $\Sigma_{N, M}$ 
can uniquely be reconstructed from the free energy $\Sigma_{1,1}$
%
\begin{tcolorbox}
${}$\\[-18pt]
\begin{equation}
\lim_{\epsilon_2 \rightarrow 0}\,\epsilon_2\, \Sigma_{N, M}(\tfrac{\rho}{N}, \cdots, \tfrac{\rho}{N}, \tfrac{\tau}{M}, \cdots, \tfrac{\tau}{M}, m, \epsilon_{1,2}) = NM \, \lim_{\epsilon_{2 \rightarrow 0}} \,\epsilon_2\,\Sigma_{1,1} (\tfrac{\rho}{N}, \tfrac{\tau}{M}, m, \epsilon_{1,2})\, .
\end{equation}
\end{tcolorbox}
\noindent We call this property of the free energies \emph{self-similarity}. Moreover, the expansion of free energies in terms of the K\"ahler parameters shows numerous recursive structures, which are generalisations of a particular Hecke-structure discovered for the case $M=1$ in the decompactification limit  $\rho\to \infty$ (which corresponds to the six-dimensional superconformal field theory limit) in \cite{Hohenegger:2015cba}.

The region (\ref{PointSimGenIntro}) in moduli space, while displaying a large enhancement of symmetry, is also significant from other perspectives. As was discussed in \cite{Hohenegger:2015cba} in the case $M=1$, when interpreting $\Sigma_{N,M}$ as counting degeneracies of monopole strings in five-dimensional supersymmetric gauge theories, the moduli space of monopole strings simplifies for this particular choice of moduli. Indeed, while in general very complicated, only at (\ref{PointSimGenIntro}) the moduli space factorises into a center-of-mass and a relative component. Thus, only in this case, the relative BPS excitations can be extracted. By S-duality of the brane configuration, a similar argument can also be applied for the cases $M\neq 1$. Furthermore, as discussed in \cite{Hohenegger:2015cba}, the modular properties of the single-particle free energies $\Sigma_{N, M} 
$
in the NS limit $\epsilon_2 \to 0$ are also better behaved at (\ref{PointSimGenIntro}).  

From the viewpoint of little string theories, the observations made above have far reaching and interesting consequences. Firstly, by construction, $\Sigma_{1,1}$ is symmetric under the exchange between $\rho$ and $\tau$, so this means that also $\Sigma_{N,M}$ for arbitrary $N, M$ acquires the same symmetry at (\ref{PointSimGenIntro}) in the NS limit.
This means that a class of ${\cal N} = (1,0)$ little string theories, those obtained by orbifolding the transverse space to type IIA or type IIB five-branes, are self-dual (\emph{i.e.} invariant) under T-duality. Secondly, the fact that the free energy $\Sigma_{N,M}$ can uniquely be reconstructed from $\Sigma_{1,1}$ indicates that the theory with arbitrary $N, M$ contains no new degrees of freedom besides (bound states of) the little strings. Finally, the emergent recursive structures indicate that the combinatorics that count the multiplicities of these bound states is simple and follows directly from geometric rules.

This paper is organised as follows: In section~\ref{Sect:LittleString} we discuss a class of little string theories with $\mathcal{N}=(1,0)$ supersymmetry that are constructed as M5-branes probing orbifold backgrounds. In section~\ref{Sect:BraneWebPart} we consider two dual descriptions of these theories: on the one hand side in terms of webs of M5-branes with stacks of M2-branes stretched in between them and on the other hand side in terms of a particular class of toric, non-compact Calabi-Yau threefolds $X_{N,M}$. We also review the BPS partition functions $\mathcal{Z}_{X_{N,M}}$ of these theories, which were first obtained in \cite{Hohenegger:2013ala}. In section~\ref{Sect:SigmaModel} we consider the partition function from the point of view of a sigma model description. The latter provides strong hints that $\mathcal{Z}_{X_{N,M}}$ factorises in the NS-limit, which indicates that the associated free energy has very interesting symmetries at (\ref{PointSimGenIntro}). In section~\ref{Sect:RelNM} we consider $\mathcal{Z}_{X_{N,M}}$ explicitly at (\ref{PointSimGenIntro}) and discuss the properties mentioned above. Section~\ref{Sect:Conclusions} contains our conclusions, while several technical points (in particular the Fourier expansions of several BPS free energies) are related to appendix~\ref{App:FreeEnergies}.

\section{Little String Theories}\label{Sect:LittleString}

In this section, we discuss different realisations of certain classes of ${\cal N}=(1,0)$ supersymmetric little string theories (LSTs): either as configurations of branes in type II string theories and M-theory or in F-theory using elliptic Calabi-Yau threefolds.

\subsection{$\mathcal{N}=(2,0)$ and $(1,1)$ Little String Theories}\label{Sect:N20LST}

There are two classes of maximally supersymmetric little string theories: a nonchiral IIa LST with ${\cal N} = (1,1)$ supersymmetry and a chiral IIb LST with ${\cal N}=(2,0)$ supersymmetry. Physically, starting from type II string theory, they are defined by taking the decoupling limit
\bea
g_{\rm st} \rightarrow 0 \qquad \mbox{while keeping } \qquad \ell_{\rm st}  = \mbox{fixed}\,,\label{LSTdecoupling}
\eea
where $g_{\rm st}$ is the string coupling constant and $\ell_{\rm st}$ is the string length.

In this, each of the two LSTs can be realised as the decoupling limit of two different setups: either of configurations of branes or geometrically of suitable orbifold backgrounds, specifically:\footnote{In this paper we only focus on LSTs of type $A$. However, it is possible to generalise the group $A_{N-1}$ to $D_N$ or $E_6, E_7, E_8$.}
\begin{itemize}
\item IIa LST of $A_{N-1}$ type: \\
LSTs with $\mathcal{N}=(1,1)$ supersymmetry can be described as the decoupling limit (\ref{LSTdecoupling}) of either a stack of $N$ NS5-branes in type IIB string theory with transverse space $\mathbb{R}^4$, or of type IIA string theory in an $A_{N-1}$ (orbifold) background.
\item IIb LST of $A_{N-1}$ type:\\
LSTs with $\mathcal{N}=(2,0)$ supersymmetry can be described as the decoupling limit (\ref{LSTdecoupling}) of either a stack of $N$ NS5-branes in type IIA string theory with transverse space $\mathbb{R}^4$, or of type IIB string theory in an $A_{N-1}$ (orbifold) background.
\end{itemize}
The realisation of the IIb LST stemming from $N$ NS5-branes in type IIA string theory also has a description in M-theory as a decoupling limit of a system of multiple coincident M5-branes where a transverse direction is compactified on $\mathbb{S}^1$ (with radius $R$). Indeed, dimensional reduction on this transverse circle relates M-theory to type IIA string theory such that the decoupling limit is
\begin{align}
&g_{st}= \frac{R}{\ell_{st}}\mapsto 0\,,&&\ell_{p}^3=R\,\ell_{st}^2\mapsto 0 \, ,  
&& \ell_{st}^2 = \frac{\ell_{p}^3}{R} =
\mbox{fixed}\, ,\label{DecouplingMtheory}
 \end{align}
where $\ell_p$ is the Planck gravity length. Here, we are taking the Planck length to zero while keeping the string length finite to get the little string theories. Taking subsequently the string length to zero, we would obtain six-dimensional field theories. \footnote{Had we taken the string length to zero first while keeping the Planck length finite, we would get supergravity coupled to five branes. Taking subsequently the Planck length to zero, we would again obtain six-dimensional field theories. Note that gravity is outright decoupled at taking the first limit in the former case, viz. the little string theories.}

BPS excitations in this system are given by M2-branes that end on the M5-branes and wrap the transverse circle. In the decoupling limit (\ref{DecouplingMtheory}),  the tension of the strings (called M-strings in \cite{Haghighat:2013gba}) corresponding to the intersection of the M2-branes with the M5-branes remains fixed
\begin{align}
R\,T_{M2}=\frac{1}{\ell_{st}^2}=\mbox{fixed}\,.
\end{align}
In order to count BPS excitations, captured by the elliptic genus of the M-string world-sheet theory \cite{Haghighat:2013gba,Hohenegger:2013ala}, the latter is compactified on $\mathbb{T}^2$ and various twists are introduced which act as regularisation.\footnote{As we mention below, the M-string world-sheet theory flows to a $\mathcal{N}=(0,2)$ sigma model with non-compact target space in the infrared. The twists act as regularisation parameters which allow us to define a finite (equivariantly regularised) elliptic genus.} Moreover, they break the world-sheet supersymmetry from $\mathcal{N}=(4,4)$ to $(0,2)$. Since the details of this brane setup are important for the remainder of this paper, we review them in detail in the following.

\subsection{M-brane Configurations}\label{Sect:BraneConfig}
\begin{figure}[h]
\vskip0.5cm
  \centering
  \includegraphics[width=10cm]{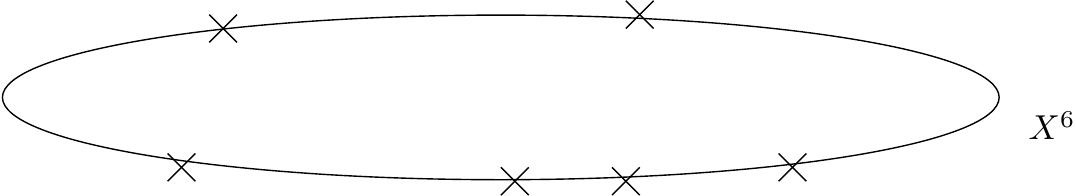}\\
  \vskip0.3cm
  \caption{\sl M5 branes separated along $x^6$.}\label{separation1}
\end{figure}

We consider M-theory on $\mathbb{R}^{5,1}\times \mathbb{S}^1\times \mathbb{R}^4_\perp$ whose coordinates are denoted by $x^{I}$ with $I=0,1,2,\ldots,10$. The M5-branes and the M2-branes are oriented according to the following table
\begin{align}
\label{braneconfig}
&\begin{array}{|c|cccccc|c|cccc|}\hline
&  x^0 & x^1& x^2 & x^3 & x^4 & \,x^5\, &\, x^6 &x^7&x^8&x^9&x^{10} \\
\hline
\text{M5-branes} & = & =  & =&=&=&=&&&&& \\
\text{M2-branes} & = & =& & & & & =&&&& \\\hline
\end{array}\\[-14pt]
&\hspace{2.2cm}\underbrace{\hspace{4.6cm}}_{\mathbb{R}^{5,1}}\, 
\hspace{-0.05cm}\underbrace{\hspace{0.8cm}}_{\mathbb{S}^1_{R_6}}\,\underbrace{\hspace{3cm}}_{\mathbb{R}_\perp^4}\nonumber
\end{align}
Here, the $N$ M5-branes are separated along the compact direction $x^6$ as shown in \figref{separation1}, while the M2-branes are stretched in between them. The M-string world-sheet theory, which extends along the directions $(x^0,x^1)$ (\emph{i.e.} the intersection between the M5- and the M2-branes), preserves $\mathcal{N}=(4,4)$ supersymmetry (see \cite{Haghighat:2013gba}).
\vskip0.5cm

\noindent
Next, we compactify the directions $(x^0,x^1)$ on $\mathbb{T}^2=\mathbb{S}^1_{R_0}\times \mathbb{S}^1_{R_1}$ and define the complex coordinates
\begin{align}
&(z^1,z^2)=(x^2+ix^3,x^4+ix^5)\,,&&\text{and} &&(w^1,w^2)=(x^7+ix^8,x^9+ix^{10})\,.\label{ComplexCords}
\end{align}
Furthermore, we introduce the following twists
\begin{itemize}
\item $\Omega$-deformation\\
Upon going around the circle $\mathbb{S}^1_{R_0}$ we twist by a $U(1)\times U(1)$ action
\begin{align}
U(1)_{\epsilon_1}\times U(1)_{\epsilon_2}:\,\,&(z_1,z_2)\rightarrow (e^{2\pi i\epsilon_1}z_1,e^{2\pi i\epsilon_2}z_2)\nonumber\\
&(w_1,w_2)\rightarrow (e^{-i\pi (\epsilon_1+\epsilon_2)}w_1,e^{-i\pi(\epsilon_1+\epsilon_2)}w_2)\,.\label{OmegaDef}
\end{align}
This deformation corresponds to the $\Omega$-background \cite{Moore:1997dj,Lossev:1997bz,Nekrasov:2002qd}.\footnote{Different realisations of the latter in string theory have been proposed in recent years \cite{Billo:2006jm,Ito:2010vx,Huang:2010kf,Huang:2011qx,Hellerman:2011mv}. In particular, a description in terms of the string world-sheet, whose starting point are physical scattering amplitudes, was proposed in \cite{Antoniadis:2010iq,Antoniadis:2013epe,Antoniadis:2013mna,Antoniadis:2015spa,Nakayama:2011be}. The connection to topological gravity has been developed in \cite{Bae:2015eoa}.}
\item mass deformation\\
Upon going around the circle $\mathbb{S}^1_{R_1}$ we twist by a $U(1)$ action
\begin{align}
U(1)_m:\,\,(w_1,w_2)\rightarrow (e^{2\pi im}w_1,e^{-2\pi im}w_2)\,,\label{MassDef}
\end{align}
\end{itemize}
Here, $m$ and $\epsilon_{1,2}$ are deformation parameters. These deformations break the world-sheet supersymmetry to $\mathcal{N}=(0,2)$.
In the infrared this world-sheet theory becomes a sigma model with target space:
\bea
\mbox{Hilb}^{k_1}[\mathbb{C}^2]\times \mbox{Hilb}^{k_2}[\mathbb{C}^2]\times \cdots \times \mbox{Hilb}^{k_N}[\mathbb{C}^2]\,,
\eea
for $k_a$ M2-branes stretched between the $a$-th and $a+1$-th M5-brane. The corresponding partition functions were studied in \cite{Haghighat:2013gba,Haghighat:2013tka,Hohenegger:2013ala}.

\subsection{$\mathcal{N}=(1,0)$ Little String Theories from Branes on Orbifolds}\label{Sect:LittleOrbifold}

In this paper we study a class of LSTs with $\mathcal{N}=(1,0)$ supersymmetry that combine features of the brane and the geometric descriptions of the $\mathcal{N}=(2,0)$ and $(1,1)$ little string theories \cite{Blum:1997mm} discussed in section~\ref{Sect:N20LST}. Specifically, these LSTs are obtained as the decoupling limit (\ref{LSTdecoupling}) of configurations of $N$ five-branes in type II string theory in the presence of an $A_{M-1}$ (orbifold) background. Depending on the details there are two different theories:
\begin{itemize}
\item $\mathbb{Z}_{N}$ orbifold of the IIa LST of $A_{M-1}$ type\\
This LST is obtained as the decoupling limit of $M$ five-branes in type IIB string theory with transverse space $\mathbb{R}^4_\perp /\mathbb{Z}_N$
\item $\mathbb{Z}_{M}$ orbifold of the IIb LST of $A_{N-1}$ type\\
This LST is obtained as the decoupling limit of $N$ five-branes in type IIA string theory with transverse space $\mathbb{R}^4_\perp /\mathbb{Z}_M$
\end{itemize}
The $\mathbb{Z}_{M}$ orbifold of the IIb LST has an alternative description as a decoupling limit of a system of multiple M5-branes probing a transverse asymptotically locally Euclidean (ALE) space of $A_{M-1}$ type (\emph{i.e.} a $\mathbb{Z}_M$ orbifold). Generalising the discussion of the un-orbifolded parent theory in section~\ref{Sect:BraneConfig}, we consider $N$ M5-branes on
\begin{align}
\mathbb{S}^1_{R_0}\times \mathbb{S}^1_{R_1}\times \mathbb{R}^4_{||}\times \mathbb{S}^1_{R_6}\times \text{ALE}_{A_{M-1}}\,,
\end{align}
which are separated along $\mathbb{S}^1_{R_6}$ (see~\figref{separation1}) and with M2-branes stretched between them. The precise setup is given in the following table \begin{align}
\label{braneconfig}
&\begin{array}{|c|c|c|cccc|c|cccc|}\hline
&  x^0 & x^1& x^2 & x^3 & x^4 & \,x^5\, &\, x^6 &x^7&x^8&x^9&x^{10} \\
\hline
\text{M5-branes} & = & =  & =&=&=&=&&&&& \\
\text{M2-branes} & = & =& & & & & =&&&& \\\hline
\end{array}\\[-14pt]
&\hspace{2.1cm}\underbrace{\hspace{0.5cm}}_{\mathbb{S}^1_{R_0}}\underbrace{\hspace{0.5cm}}_{\mathbb{S}^1_{R_1}}\underbrace{\hspace{3cm}}_{\mathbb{R}^{4}_{||}}\, 
\hspace{-0.05cm}\underbrace{\hspace{0.5cm}}_{\mathbb{S}^1_{R_6}}\,\underbrace{\hspace{3cm}}_{\text{ALE}_{A_{M-1}}}\nonumber
\end{align}
The configuration is also depicted in Fig. 2.
\begin{figure}[h]
  \centering
  \includegraphics[width=17cm]{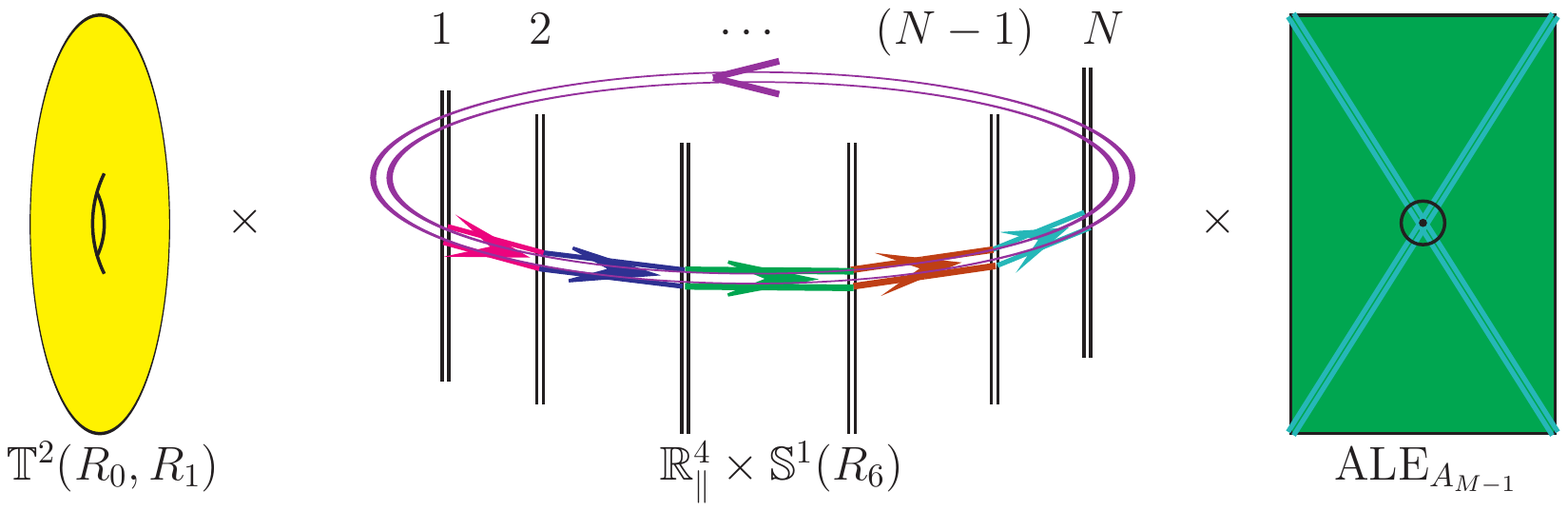}\\
  \vskip-17cm
  \caption{\sl The M-brane configuration for orbifolded little string theories.}\label{brane-configuration}
  \vskip1cm
\end{figure}
\noindent
Here, the orthogonal ALE space is defined as an orbifold of $\mathbb{R}^4_\perp$, specifically
\begin{align}
&\text{ALE}_{A_{M-1} }:= \mathbb{R}^4_{\perp}/ \mathbb{Z}_M\,,&&\text{with}&&
\mathbb{Z}_M = \left\{ \begin{pmatrix} e^{\frac{2\pi i\,n}{M}}& 0\\ 0 & e^{-\frac{2\pi i\,n}{M}} \end{pmatrix}, \, n = 0, \cdots, M-1 \right\}\,,\label{DefOrbALE}
\end{align}
where $\mathbb{Z}_M$ acts on the complex coordinates $(w_1,w_2)$ as  defined in (\ref{ComplexCords}). The M5-branes are transverse to the $\text{ALE}_{A_{M-1} }$ space and are located at the singularity of the orbifold.

As mentioned in section~\ref{Sect:BraneConfig}, in the un-orbifolded case (\emph{i.e.} for $M=1$), the M-string world-sheet theory preserves $\mathcal{N}=(0,2)$ supersymmetry. The supercharges transform in the representation $({\bf 1},{\bf 1},{\bf 1},{\bf 2})_{-\frac{1}{2}}$ of $Spin(4)_{\parallel}\times Spin(4)_{\perp}\times Spin(1,1)$, which is the bosonic part of the superconformal group (see \cite{Haghighat:2013gba} for further details). Furthermore, the orbifold (\ref{DefOrbALE}) only acts on the $SU(2)_{L}$ subgroup of $Spin(4)_{\perp}=SU(2)_{L}\times SU(2)_{R}$. 
The supercharges are singlets under the $SU(2)_L$, so they are preserved by the orbifold action. Thus, even after orbifolding, the world-sheet theory still preserves $(0,2)$  supersymmetry and flows to a $\mathcal{N}=(0,2)$ sigma model in the infrared, whose target space is\footnote{We will discuss this sigma model in further detail in section~\ref{Sect:SigmaModel}.}
\begin{align}
\mathfrak{M}_{N,k_{1}}\times \mathfrak{M}_{N,k_{2}}\times \cdots\times \mathfrak{M}_{N,k_{M}}\, ,
\end{align}
for $k_a$ M2-branes stretched between the $a$-th and $a+1$-th M5-brane. Here, $\mathfrak{M}_{N,k}$ is the moduli space for $U(N)$ instantons of charge $k$.

In order to define the BPS counting functions for the M-string theory, we need to introduce the $\Omega$- and mass-deformations as in section~\ref{Sect:BraneConfig}. As discussed in \cite{Haghighat:2013tka}, these deformations can also be introduced when $\mathbb{R}^4_{\perp}$ is replaced by $\text{ALE}_{A_{M-1}}$. Indeed, geometrically, the latter can be understood as an $\mathbb{S}^1$ fibration over $\mathbb{R}^3$ and has two isometries associated with the fiber and the base, respectively,
\bea
&& U(1)_F: \qquad (w_1, w_2) \quad \rightarrow \quad (e^{+2 \pi i \mu} w_1, e^{-2 \pi i \mu} w_2) \,,\nonumber \\
&& U(1)_B: \qquad (w_1, w_2) \quad \rightarrow \quad (e^{+2 \pi i \nu} w_1, e^{+2 \pi i \nu} w_2)\,.
\eea
Here, $(w_1,w_2)$ have been introduced in (\ref{ComplexCords}) and $\mu$ and $\nu$ are arbitrary real parameters. These isometries allow us to introduce the $\Omega$- and mass deformations in the same manner as in (\ref{OmegaDef}) and (\ref{MassDef}), respectively.

\subsection{Little String Theories from F-Theory}\label{Sect:Ftheory}

In this section we recapitulate briefly the construction of little string theories from F-theory involving elliptic Calabi-Yau threefolds \cite{Morrison:1996na,Morrison:1996pp}. For a more detailed discussion, including a classification of little string theories from F-theory, see \cite{Bhardwaj:2015oru}.

As we discussed in section~\ref{Sect:N20LST}, the $(2,0)$ little string theories of $A$ type are realized in type IIB by compactification on an $A_{N-1}$ orbifold. We can resolve the singularities of this space by replacing them with $N$ $\mathbb{P}^1$'s. However, in order to have a compact moduli space, the size of the latter should not be arbitrary: indeed, one possibility is to consider an affine orbifold, \emph{i.e.} an elliptic fibration with a singular fiber of type $I_{N-1}$ \cite{kodaira}, which we will denote by $\text{ALE}_{\widehat{A}_{N-1}}$. In this case, the size of the elliptic fiber sets the scale for the size of the resolved cycles. Each cycle in the resolution corresponds to a linear combination of roots of the $\widehat A_{N-1}$ algebra. The elliptic fiber corresponds to the imaginary root 
\bea
\delta=\alpha_{1}+\alpha_{2}+\cdots +\alpha_{N}
\eea
where $\alpha_{i}$ are the real simple roots of $\widehat A_{N-1}$. Translated to the K\"ahler parameters, we get
\bea
\rho=t_{1}+t_{2}+\cdots+t_{N}\,,
\eea
where $\rho$ is the K\"ahler parameter of the elliptic curve and $t_{i}$ are the K\"ahler parameters of the $N$ $\mathbb{P}^1$'s. In the lift to F-theory, the compactification space becomes $\mathbb{T}^2$ fibered over $\text{ALE}_{\widehat{A}_{N-1}}$. This space has $\mathbb{T}^2\times \mathbb{T}^2$ fibration over the complex plane, where one of the elliptic fibrations is a singular fiber of type $I_{N-1}$.

Generalising this description of the $(2,0)$ little strings in terms of Calabi-Yau threefolds with a double elliptic fibration, we can consider an elliptic Calabi-Yau threefold in which one elliptic fibration develops a singularity of type $I_{N-1}$ and the other develops a singularity of type $I_{M-1}$. This more general situation corresponds to a class of ${\cal N}=(1,0)$ little string theories \cite{Blum:1997mm} that can be constructed from type IIB string theory by introducing $M$ NS5-branes transverse to  $\text{ALE}_{\widehat A_{N-1}}$. These Calabi-Yau threefolds are toric and have dual web diagrams which were studied in \cite{Hohenegger:2013ala}. Using the topological vertex formalism, the corresponding refined topological string partition functions and their modular properties were studied in \cite{Hohenegger:2013ala,Hohenegger:2015btj}.
If we take the singular fibers of either of the two elliptic fibrations to be the base, then the total space is either an  $\text{ALE}_{\widehat A_{N-1}}$ space fibered over a circular chain of $M$ $\mathbb{P}^1$'s or as an $\text{ALE}_{\widehat A_{M-1}}$ space fibered over a circular chain of $N$ $\mathbb{P}^1$'s.
These two descriptions, studied in \cite{Hohenegger:2015btj}, correspond to two different gauge theories and can be understood in terms of fiber-base duality.

\section{Brane Webs and Topological Amplitudes}\label{Sect:BraneWebPart}
\subsection{From Little Strings to Topological Strings}

In the last section, we have seen that little strings in orbifold backgrounds are dual to Calabi-Yau threefolds with a double elliptic fibration structure \cite{Bhardwaj:2015oru}. For A-type orbifolds, the corresponding Calabi-Yau threefolds are dual to type IIB $(N,M)$-brane webs \cite{Leung:1997tw} as well.  We can calculate the little string partition function from the topological string partition function of this Calabi-Yau threefold, as to be discussed in detail in this section.

\begin{figure}[htb]
\begin{center}
\begin{tikzpicture}[scale = 0.70]
\draw[ultra thick,green!50!black] (-6,0) -- (-5,0);
\draw[ultra thick,red] (-5,-1) -- (-5,0);
\draw[ultra thick,blue] (-5,0) -- (-4,1);
\draw[ultra thick,green!50!black] (-4,1) -- (-3,1);
\draw[ultra thick,red] (-4,1) -- (-4,2);
\draw[ultra thick,red] (-3,1) -- (-3,0);
\draw[ultra thick,blue] (-3,1) -- (-2,2);
\draw[ultra thick,blue] (-4,2) -- (-3,3);
\draw[ultra thick,green!50!black] (-5,2) -- (-4,2);
\draw[ultra thick,red] (-3,3) -- (-3,4);
\draw[ultra thick,green!50!black] (-3,3) -- (-2,3);
\draw[ultra thick,red] (-2,2) -- (-2,3);
\draw[ultra thick,green!50!black] (-2,2) -- (-1,2);
\draw[ultra thick,blue] (-2,3) -- (-1,4);
\draw[ultra thick,blue] (-1,2) -- (0,3);
\draw[ultra thick,red] (-1,2) -- (-1,1);
\draw[ultra thick,green!50!black] (-4,4) -- (-3,4);
\draw[ultra thick,blue] (-3,4) -- (-2,5);
\draw[ultra thick,green!50!black] (-2,5) -- (-1,5);
\draw[ultra thick,green!50!black] (-1,4) -- (0,4);
\draw[ultra thick,green!50!black] (0,3) -- (1,3);
\draw[ultra thick,red] (-2,5) -- (-2,6);
\draw[ultra thick,red] (-1,4) -- (-1,5);
\draw[ultra thick,red] (0,3) -- (0,4);
\draw[ultra thick,blue] (1,3) -- (2,4);
\draw[ultra thick,blue] (0,4) -- (1,5);
\draw[ultra thick,blue] (-1,5) -- (0,6);
%
\draw[ultra thick,green!50!black] (2,4) -- (3,4);
\draw[ultra thick,green!50!black] (1,5) -- (2,5);
\draw[ultra thick,green!50!black] (0,6) -- (1,6);
%
\draw[ultra thick,red] (1,3) -- (1,2);
\draw[ultra thick,red] (2,4) -- (2,5);
\draw[ultra thick,red] (1,5) -- (1,6);
\draw[ultra thick,red] (0,6) -- (0,7);
\draw[ultra thick,green!50!black] (-1,10) -- (0,10);
\node[rotate=90] at (0,8) {{\Huge$\ldots$}};
\node[rotate=90] at (2,9) {{\Huge$\ldots$}};
\node[rotate=90] at (4,10) {{\Huge$\ldots$}};
\node[rotate=90] at (6,11) {{\Huge$\ldots$}};
\node[rotate=90] at (-2,7) {{\Huge$\ldots$}};
\draw[ultra thick,red] (0,9) -- (0,10);
\draw[ultra thick,blue] (0,10) -- (1,11);
\draw[ultra thick,red] (1,11) -- (1,12);
\draw[ultra thick,red] (2,11) -- (2,10);
\draw[ultra thick,green!50!black] (1,11) -- (2,11);
\draw[ultra thick,blue] (2,11) -- (3,12);
\draw[ultra thick,red] (3,12) -- (3,13);
\draw[ultra thick,red] (4,12) -- (4,11);
\draw[ultra thick,green!50!black] (3,12) -- (4,12);
\draw[ultra thick,blue] (4,12) -- (5,13);
\draw[ultra thick,green!50!black] (5,13) -- (6,13);
\draw[ultra thick,red] (5,13) -- (5,14);
\draw[ultra thick,blue] (6,13) -- (7,14);
\draw[ultra thick,red] (6,13) -- (6,12);
\draw[ultra thick,green!50!black] (7,14) -- (8,14);
\draw[ultra thick,red] (7,14) -- (7,15);
\draw[ultra thick,blue] (1,6) -- (2,7);
\draw[ultra thick,red] (2,7) -- (2,8);
\draw[ultra thick,red] (3,7) -- (3,6);
\draw[ultra thick,green!50!black] (2,7) -- (3,7);
\draw[ultra thick,blue] (3,7) -- (4,8);
\draw[ultra thick,red] (4,8) -- (4,9);
\draw[ultra thick,red] (5,8) -- (5,7);
\draw[ultra thick,green!50!black] (4,8) -- (5,8);
\draw[ultra thick,green!50!black] (3,6) -- (4,6);
\draw[ultra thick,green!50!black] (5,7) -- (6,7);
\draw[ultra thick,blue] (5,8) -- (6,9);
\draw[ultra thick,red] (6,9) -- (6,10);
\draw[ultra thick,blue] (2,5) -- (3,6) ;
\draw[ultra thick,blue] (4,6) -- (5,7) ;
\draw[ultra thick,green!50!black] (6,9) -- (7,9);
\node at (8,9) {{\Huge $\ldots$}};
\draw[ultra thick,blue] (3,4) -- (4,5);
\draw[ultra thick,red] (3,4) -- (3,3);
\draw[ultra thick,red] (4,5) -- (4,6);
\draw[ultra thick,green!50!black] (4,5) -- (5,5);
\node at (7,7) {{\Huge $\ldots$}};
\draw[ultra thick,green!50!black] (10,14) -- (11,14);
\draw[ultra thick,red] (11,14) -- (11,13);
\draw[ultra thick,blue] (11,14) -- (12,15);
\node at (9,14) {{\Huge $\ldots$}};
\draw[ultra thick,red] (12,15) -- (12,16);
\draw[ultra thick,green!50!black] (12,15) -- (13,15);

\node[rotate=90] at (11,12) {{\Huge $\ldots$}};
\node at (6,5) {{\Huge $\ldots$}};
\draw[ultra thick,blue] (10,9) -- (11,10);
\draw[ultra thick,green!50!black] (9,9) -- (10,9);
\draw[ultra thick,red] (11,10) -- (11,11);
\draw[ultra thick,green!50!black] (11,10) -- (12,10);
\draw[ultra thick,red] (10,9) -- (10,8);
\draw[ultra thick,blue] (9,7) -- (10,8);
\draw[ultra thick,green!50!black] (8,7) -- (9,7);
\draw[ultra thick,red] (9,6) -- (9,7);
\draw[ultra thick,green!50!black] (10,8) -- (11,8);
\draw[ultra thick,blue] (8,5) -- (9,6);
\draw[ultra thick,green!50!black] (9,6) -- (10,6);
\draw[ultra thick,green!50!black] (7,5) -- (8,5);
\draw[ultra thick,red] (8,4) -- (8,5);
%
%
%
\draw[ultra thick,green!50!black] (-3,9) -- (-2,9);
\draw[ultra thick,blue] (-2,9) -- (-1,10);
\draw[ultra thick,red] (-1,10) -- (-1,11);
\draw[ultra thick,red] (-2,8) -- (-2,9);
\node[rotate=90] at (-5.5,0) {$=$};
\node at (-5.3,-0.2) {{\tiny$1$}};
\node[rotate=90] at (-4.5,2) {$=$};
\node at (-4.3,1.8) {{\tiny$2$}};
\node[rotate=90] at (-3.5,4) {$=$};
\node at (-3.3,3.8) {{\tiny$3$}};
\node[rotate=90] at (-2.5,9) {$=$};
\node at (-2.3,8.75) {{\tiny$M$}};
\node[rotate=90] at (9.5,6) {$=$};
\node at (9.7,5.8) {{\tiny$1$}};
\node[rotate=90] at (10.5,8) {$=$};
\node at (10.7,7.8) {{\tiny$2$}};
\node[rotate=90] at (11.5,10) {$=$};
\node at (11.7,9.8) {{\tiny$3$}};
\node[rotate=90] at (12.7,15) {$=$};
\node at (12.9,14.75) {{\tiny$M$}};
\node at (-5,-0.5) {$-$};
\node at (-4.8,-0.7) {{\tiny $1$}};
\node at (-3,0.5) {$-$};
\node at (-2.8,0.3) {{\tiny $2$}};
\node at (-1,1.5) {$-$};
\node at (-0.8,1.3) {{\tiny $3$}};
\node at (1,2.5) {$-$};
\node at (1.2,2.3) {{\tiny $4$}};
\node at (3,3.5) {$-$};
\node at (3.2,3.3) {{\tiny $5$}};
\node at (8,4.5) {$-$};
\node at (8.25,4.3) {{\tiny $N$}};
\node at (-1,10.5) {$-$};
\node at (-0.8,10.3) {{\tiny $1$}};
\node at (1,11.5) {$-$};
\node at (1.2,11.3) {{\tiny $2$}};
\node at (3,12.5) {$-$};
\node at (3.2,12.3) {{\tiny $3$}};
\node at (5,13.5) {$-$};
\node at (5.2,13.3) {{\tiny $4$}};
\node at (7,14.5) {$-$};
\node at (7.2,14.3) {{\tiny $5$}};
\node at (12,15.5) {$-$};
\node at (12.25,15.3) {{\tiny $N$}};
\draw[thick, <->] (-5,-2) -- (-3.05,-2);
\node at (-4,-2.5) {$t_1$};
\draw[thick, <->] (-2.95,-2) -- (-1.05,-2);
\node at (-2,-2.5) {$t_2$};
\draw[thick, <->] (-0.95,-2) -- (0.95,-2);
\node at (0,-2.5) {$t_3$};
\draw[thick, <->] (1.05,-2) -- (2.95,-2);
\node at (2,-2.5) {$t_4$};
\draw[thick, <->] (3.05,-2) -- (4.95,-2);
\node at (4,-2.5) {$t_5$};
\node at (6.5,-2) {{\Huge $\ldots$}};
\draw[thick, <->] (8.05,-2) -- (9.95,-2);
\node at (9,-2.5) {$t_N$};
\draw[thick, <->] (-7,0) -- (-7,1.95);
\node at (-7.5,1) {$T_1$};
\draw[thick, <->] (-7,2.05) -- (-7,3.95);
\node at (-7.5,3) {$T_2$};
\node[rotate=90] at (-7,6.5) {{\Huge $\ldots$ }};
\draw[thick, <->] (-7,9.05) -- (-7,10.95);
\node at (-7.5,10) {$T_M$};
\end{tikzpicture}
\end{center}
\caption{\sl The $(N, M)$-brane web diagram. It is also the toric diagram of the Calabi-Yau threefold $X_{N,M}$. The length of the blue lines represents the mass-deformation $m$, which is considered to be the same throughout the whole diagram.}
\label{Fig:WebToric}
\end{figure}
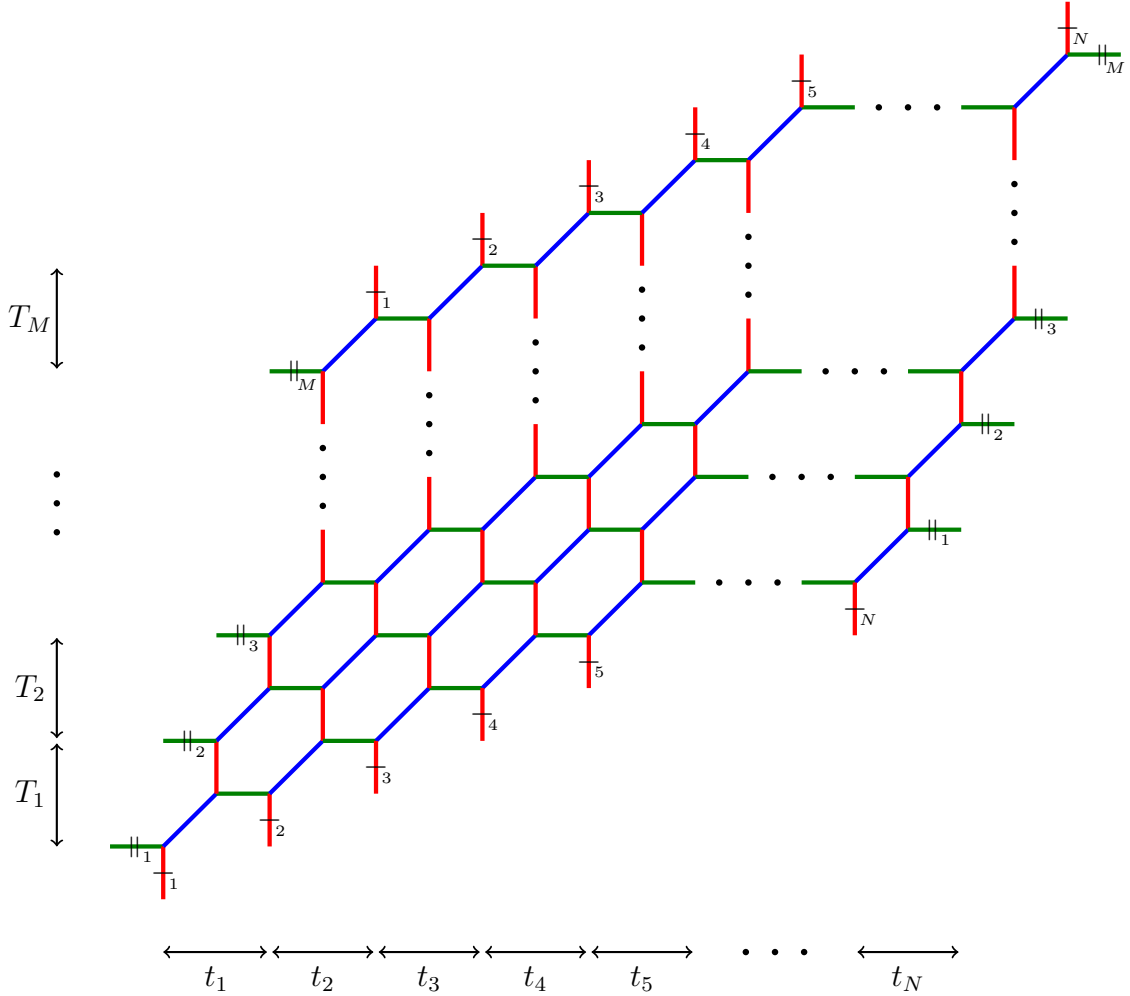

 We denote by $X_{N,M}$ the Calabi-Yau threefolds dual to the $(N,M)$-brane web shown in \figref{Fig:WebToric}. The $X_{N,M}$ are double elliptic fibrations over a non-compact base and can also be thought of as resolutions of $\mathbb{Z}_{N}\times \mathbb{Z}_{M}$ orbifolds of $X_{1,1}$. At the boundary of K\"ahler parameter space, $X_{1,1}$ becomes the resolved conifold, whose orbifolds were also studied previously (see, for instance, \cite{Aganagic:1999fe}).


We write the refined topological string partition function of $X_{N,M}$ as
\begin{align}
{\cal Z}_{X_{N,M}}(\tau,\rho&,t_{1},\ldots, t_{N-1},T_{1},\ldots,T_{M-1},m,\epsilon_{1,2})\,.\label{DefTopPartFctMN}
\end{align}
The function ${\cal Z}_{X_{N,M}}$ depends on the parameters of the brane web, which are the K\"ahler parameters of $X_{N,M}$. The number of parameters in the most generic case can be counted from \figref{Fig:WebToric}, as discussed in \cite{Haghighat:2013gba}. Here, we review this counting for completeness: If the web in \figref{Fig:WebToric} is not compactified (both horizontally and vertically), there are $(N-1)$ parameters from the separation between the vertical NS5-branes and $(M-1)$ parameters from the separation between the horizontal D5-branes. Since these $N$ NS5-branes and $M$ D5-branes intersect at $M\,N$ points in the $(x^5,x^6)$-plane, which can be resolved independently, we have a total of $M\,N+(N-1)+(M-1)$ moduli. Upon compactification of the web in the vertical direction, the separation between neighbouring vertical branes at the bottom of the diagram must be the same as the separation between neighbouring vertical branes at the top. For example, using the notation of \figref{Fig:WebToric}, the separation between the two red lines labelled by $|_1$ and $|_2$ must be the same at the bottom and the top of the web. This places $(N-1)$ constraints and therefore the number of moduli is $M\,N+(M-1)+1$, where we have included one more moduli which is the radius $\tau$ of the circle of compactification. Upon compactifying also the horizontal direction we have similar constraints for the horizontal branes, \emph{e.g.} the separation between the green lines marked as $||_1$ and $||_2$ in \figref{Fig:WebToric} must be the same on the left and the right side of the web. This places additional $(M-1)$ constraints such that the total number of moduli is
\bea
\underbrace{M\,N}_{\text{intersections}}+\underbrace{(N-1)}_{T_{a}}+\underbrace{(N-1)}_{t_i}-\underbrace{(M-1)}_{\text{horizontal constraint}}-\underbrace{(N-1)}_{\text{vertical constraint}}+\underbrace{2}_{\tau,\rho}=M\,N+2
\eea
where we added a further modulus corresponding to the radius of the horizontal circle $\rho$.

In this paper (for the partition function (\ref{DefTopPartFctMN})) we are not considering the most general case: for simplicity, we choose to resolve the intersection points by associating the same parameter $m$ with each of them. Thus, we have only $1$ rather than $M\,N$ moduli coming from the resolution of the intersections. In this case, however, the constraints coming from the compactification on $\mathbb{S}^1_{R_1}\times \mathbb{S}^1_{R_6}$, as described above, are automatically satisfied since by construction the separation between the branes is the same on either side. Therefore, the total number of moduli is:
\bea
\underbrace{1}_{m}+\underbrace{(M-1)}_{T_{a}}+\underbrace{(N-1)}_{t_{i}}+\underbrace{2}_{\tau,\rho}=M+N+1\,.
\eea
Explicitly, apart from $m$, the remaining $M+N$ parameters can be chosen to be
\begin{align}
&T_{i}=\mathbf{b}_{i+1}-\mathbf{b}_i\,,&&\forall i=1,\ldots, M\,,\nonumber\\
&t_{a}=\mathbf{a}_{a+1}-\mathbf{a}_a\,,&&\forall a=1,\ldots, N\,,
\end{align}
where $\mathbf{a}_a$ and $\mathbf{b}_i$ are the positions of the five-branes on the circles $\mathbb{S}^1_{R_1}\times \mathbb{S}^1_{R_6}$, along the $x^1$ and $x^6$ directions respectively. We can also introduce the quantities $\tau$ and $\rho$ through
\begin{align}
&\tau=\frac{i}{2\pi}(T_{1}+T_{2}+\cdots+T_{M})\,,&&\rho=\frac{i}{2\pi}(t_{1}+t_{2}+\cdots+t_{N})\,,
\end{align}
which are the K\"ahler parameters associated with the two elliptic fibers of $X_{N,M}$. In the explicit form of (\ref{DefTopPartFctMN}) these variables will appear in the following combinations
{\allowdisplaybreaks\begin{align}
&\overline{Q}_{i}=e^{- T_{i}}\,,&&\forall i=1,\ldots,M\,,\nonumber\\
&Q_{a}=e^{- t_{a}}\,,&&\forall a=1,\ldots,N\,,\nonumber\\
&Q_{ab}=Q_aQ_{a+1}\ldots Q_{b-1}\,,&& 1\leq a<b\leq N\,,
\end{align}}
as well as
\begin{align}
&Q_{\tau}=e^{2\pi i \tau}\,,&&Q_{\rho}=e^{2\pi i \rho}\,.
\end{align}
In order to save writing we introduce the following notation for the collection of all parameters
\begin{align}
&\mathbf{T}=(T_1,T_2\ldots,T_M)\,,&&\text{and}&&\mathbf{t}=(t_1,t_2,\ldots,t_N)\,.
\end{align}
The refined topological string partition function $\mathcal{Z}_{X_{N,M}}$ for the class of Calabi-Yau threefolds $X_{N,M}$ was calculated in \cite{Hohenegger:2013ala} using the refined topological vertex and we refer the reader to that paper for more details. Depending on the choice of the preferred direction of the vertex -- vertical or horizontal -- it is given by two different looking (yet equivalent) expressions.

\subsection{Topological String Partition Function}\label{Sect:PartFct}
In this section we discuss in detail the partition function ${\cal Z}_{X_{N,M}}$ (computed with the help of the refined topological vertex) and the two equivalent forms in which it can be written depending on the choice of the preferred direction of the vertex.

Explicitly, the partition function ${\cal Z}_{X_{N,M}}$ takes the following form \cite{Hohenegger:2013ala}:
\begin{align}
{\cal Z}_{X_{N,M}}&(\tau,\rho,t_{1},\ldots, t_{N-1},T_{1},\ldots,T_{M-1},m,\epsilon_{1,2})=W_N(\emptyset)^M\,\sum_{\alpha_a^{(i)}}Q_\tau ^{\sum_a|\alpha^{(M)}_a|}\,\left(\prod_{i=1}^M \overline{Q}_i^{\sum_a\left(\left|\alpha_a^{(i)}\right|-\left|\alpha_a^{(M)}\right|\right)}\right)\nonumber\\
&\times \prod_{i=1}^M\prod_{a=1}^N\frac{\vartheta_{\alpha_a^{(i+1)}\alpha_a^{(i)}}(Q_m;\rho)}{\vartheta_{\alpha_a^{(i)}\alpha_a^{(i)}}(\sqrt{t/q};\rho)}\,\prod_{1\leq a<b\leq N}\prod_{i=1}^M\frac{\vartheta_{\alpha_a^{(i)}\alpha_b^{(i+1)}}(Q_{ab}Q_m^{-1};\rho)\,\vartheta_{\alpha_a^{(i+1)}\alpha_b^{(i)}}(Q_{ab}Q_m;\rho)}{\vartheta_{\alpha_a^{(i)}\alpha_b^{(i)}}(Q_{ab}\sqrt{t/q};\rho)\,\vartheta_{\alpha_a^{(i)}\alpha_b^{(i)}}(Q_{ab}\sqrt{q/t};\rho)}\,,\label{FunctTopPartFctMN}
\end{align}
with
\begin{align}
&W_N(\emptyset;t_1,\ldots,t_N,m,\epsilon_{1,2})=\lim_{\tau\to i\infty}{\cal Z}_{X_{N,1}}(\tau,\rho,t_{1},\cdots,t_{N-1},m,\epsilon_{1,2})\nonumber\\
&=\frac{1}{\prod_{n=1}^\infty(1-Q_\rho^n)}\prod_{i,j,k}\prod_{a,b=1}^N\frac{1-Q_\rho^{k-1}Q_{a,a+b}Q_m^{-1}t^{i-\frac{1}{2}}q^{j-\frac{1}{2}}}{1-Q_\rho^{k-1}Q_{a,a+b}\,t^{i-1}q^{j}}\,\frac{1-Q_\rho^{k-1}Q_{a,a+b-1}Q_m\, t^{i-\frac{1}{2}}q^{j-\frac{1}{2}}}{1-Q_\rho^{k-1}Q_{a,a+b}\,t^{i}q^{j-1}}\,.\label{DefWNfact}
\end{align}
The summation in (\ref{FunctTopPartFctMN}) is over $N M$ independent integer partitions $\alpha_a^{(i)}$ and for a specific $\alpha_a^{(i)}=(\alpha_{a,1}^{(i)},\alpha_{a,2}^{(i)},\ldots,\alpha_{a,\ell}^{(i)})$ of length $\ell$, we have introduced
\begin{align}
|\alpha_{a}^{(i)}|=\sum_{n=1}^\ell\alpha_{a,n}^{(i)}\,.
\end{align}
Furthermore, given two integer partitions $\mu$ and $\nu$, we have used the following definitions in (\ref{FunctTopPartFctMN})
\begin{align}
\vartheta_{\mu\nu}(x;\rho):=\prod_{(i,j)\in\mu}\vartheta\left(x^{-1}\,q^{-\mu_i+j-\tfrac{1}{2}}\,t^{-\nu_j^t+i-\tfrac{1}{2}};\rho\right)\,\prod_{(i,j)\in\nu}\vartheta\left(x^{-1}\,q^{\nu_i-j+\tfrac{1}{2}}\,t^{\mu_j^t-i+\tfrac{1}{2}};\rho\right)\,,
\end{align}
where (for $x=e^{2\pi iz}$)
\begin{align}
\vartheta(x;\rho):=\left(x^{\frac{1}{2}}-x^{-\frac{1}{2}}\right)\prod_{k=1}^{\infty}(1-x\,Q_\rho^k)(1-x^{-1}\,Q_\rho^k)=\frac{i\,\theta_1(\rho,x)}{Q_\rho^{1/8}\,\prod_{k=1}^\infty(1-Q_\rho^k)}\,.\label{DefVarthet}
\end{align}
The spin content of the BPS states is encoded in the free energy and using the plethystic logarithm of the refined partition function we can obtain the BPS counting function for the single particle states:
\begin{align}\nonumber
\Sigma_{N,M}({\bf t},{\bf T},m,\epsilon_{1,2})&=\text{Plog}\,{\cal Z}_{X_{N,M}}(\tau,\rho,t_1,\ldots,t_{N-1},T_1,\ldots,T_{M-1},m,\epsilon_{1,2})\\\nonumber
&=\sum_{k=1}^{\infty}\frac{\mu(k)}{k}\,\text{log}{\cal Z}_{X_{N,M}}(k\tau,k\rho,k\,t_1,\ldots,k\,t_{N-1},k\,T_1,\ldots,k\,T_{M-1},k\,m,k\epsilon_{1,2})\,,
\end{align}
where $\mu(k)$ is the M\"obius function. We can write $\Sigma_{N,M}$ as the following power series in $\overline{Q}_{i}$
\bea
\Sigma_{N,M}({\bf t}, {\bf T},m,\epsilon_{1,2})&=&\sum_{k_1,\ldots,k_M=0}^\infty\overline{Q}_1^{k_1}\ldots\overline{Q}_M^{k_M}\,\Sigma_{N,M}^{(k_1,\ldots, k_M)}({\bf t},m,\epsilon_{1,2})\,\nonumber\\
&=&\sum_{r_1,\ldots,r_N=0}^\infty Q_1^{r_1}\ldots Q_N^{r_N}\,G_{N,M}^{(r_1,\ldots, r_N)}({\bf T},m,\epsilon_{1,2})\,.
\eea
Following the notation of \cite{Hohenegger:2015btj}, we will refer to the expansion in terms of $\overline{Q}_i$ (with free energies $\Sigma_{N,M}^{(\{k_i\})}$) as \emph{vertical} and the expansion in terms of $Q_a$ (with free energies $G_{N,M}^{(\{r_a\})}$) as \emph{horizontal}. As mentioned above, this notation refers to the preferred direction of the (refined) topological vertex when computing the topological string partition function.\footnote{We note that in \cite{Hohenegger:2015btj} the vertical and horizontal partition function have been normalised in slightly different ways. Here we keep the normalisation of both expansions to be the same.}

While the free energies $\Sigma_{N,M}$ are in general rather complicated, in the following we will be interested in the limit\footnote{In (\ref{limit}) we consider $\tau$ and $\rho$ as fixed (albeit arbitrary) parameters. It is therefore that we refer to (\ref{limit}) as a 'point' in the moduli space.}
\begin{align}
Q_{1}=Q_{2}=\ldots=Q_{N}=Q_{\rho}^{\frac{1}{N}}&&\text{and} &&
\overline{Q}_{1}=\overline{Q}_{2}=\ldots =\overline{Q}_{M}=Q_{\tau}^{\frac{1}{M}}\,.\label{limit}
\end{align}
as well as the NS-limit $\epsilon_2\to 0$. In order to describe the latter, we introduce\footnote{As a function of the $\epsilon$-deformations $\Sigma_{N,M}({\bf t},{\bf T},m,\epsilon_{1,2})$ diverges as $\frac{1}{\epsilon_{1}\epsilon_{2}}$ in the limit $\epsilon_{1,2}\mapsto 0$. This explains the additional factor of $\epsilon_2$ in (\ref{VertFreeEnergies}).}
\begin{align}
\widetilde{\Sigma}_{N,M}^{(K)}({\bf t},m,\epsilon_1)=\lim_{\epsilon_2\to 0}\,\epsilon_2\,\sum_{\{k_i\},\sum k_i=K}\,\Sigma_{N,M}^{(k_1,\ldots, k_M)}({\bf t},m,\epsilon_{1,2})\,,\label{VertFreeEnergies}
\end{align}
where the sum is over all configurations $(k_1,\ldots, k_M)$ such that $\sum_{i=1}^Mk_i=K\in\mathbb{N}$. In this way we have
\begin{align}
\lim_{\epsilon_2\to 0}\epsilon_2\Sigma_{N,M}\left(\tfrac{\rho}{N},\ldots, \tfrac{\rho}{N},\tfrac{\tau}{M},\ldots,\tfrac{\tau}{M},m,\epsilon_{1,2}\right)&=\sum_{K=0}^\infty Q_\tau^{\frac{K}{M}}\,\widetilde{\Sigma}_{N,M}^{(K)}\left(\tfrac{\rho}{N},m,\epsilon_1\right)\,,\label{FreeEnergyNS}\\\nonumber
&=\sum_{R=0}^\infty Q_\rho^{\frac{R}{N}}\,\widetilde{G}_{N,M}^{(R)}\left(\tfrac{\tau}{M},m,\epsilon_1\right)\,.
\end{align}
where by abuse of notation we have written $\widetilde{\Sigma}_{N,M}^{(K)}\left(\tfrac{\rho}{N},m,\epsilon_1\right)=\widetilde{\Sigma}_{N,M}^{(K)}\left(\tfrac{\rho}{N},\ldots,\tfrac{\rho}{N},m,\epsilon_1\right)$ as well as $\widetilde{G}_{N,M}^{(R)}\left(\tfrac{\tau}{M},m,\epsilon_1\right)=\widetilde{G}_{N,M}^{(R)}\left(\tfrac{\tau}{M},\ldots,\tfrac{\tau}{M},m,\epsilon_1\right)$.

\subsection{Little String Partition Function and T-Duality}
As we have mentioned in section~\ref{Sect:Ftheory}, the orbifold backgrounds of the little strings discussed in section~\ref{Sect:LittleOrbifold} are related to the toric, non-compact Calabi-Yau threefolds through a chain of dualities. Therefore, also the BPS degeneracies of the little strings can be extracted from the topological string on $X_{N,M}$. Specifically, the partition function of the $\mathbb{Z}_M$ orbifold of the type IIa LST on $A_{N-1}$ (denoted as $Z^{(N,M)}_{\text{IIa}}$ in the following) and the partition function of the $\mathbb{Z}_N$ orbifold of the IIb LST on $A_{M-1}$ (denoted as $Z^{(M,N)}_{\text{IIb}}$ in the following) can be expressed in terms of $\mathcal{Z}_{X_{N,M}}$ introduced in (\ref{FunctTopPartFctMN}) in the following manner
\bea
Z^{(N,M)}_{\text{IIa}}(\mathbf{T},\mathbf{t},m,\epsilon_{1,2})={\cal Z}_{X_{N,M}}(\tau,\rho,t_{1},\ldots, t_{N-1},T_{1},\ldots,T_{M-1},m,\epsilon_{1,2})\,,\\
Z^{(M,N)}_{\text{IIb}}(\mathbf{T},\mathbf{t},m,\epsilon_{1,2})={\cal Z}_{X_{M,N}}(\rho,\tau,T_{1},\ldots, T_{M-1},t_{1},\ldots,t_{N-1},m,\epsilon_{1,2})\,.
\eea
For $Z_{\rm IIa}$, the first set of arguments $\mathbf{T}$ denotes the K\"ahler parameters of the $\mathbb{Z}_M$ orbifold while the second set of arguments $\mathbf{t}$ denotes separations of NS5-branes. For $Z_{\rm IIb}$, the first set of arguments $\mathbf{T}$ denotes separations of NS5-branes while the second set of arguments $\mathbf{t}$ denotes the K\"ahler parameters of the $\mathbb{Z}_N$ orbifold.
From the perspective of the little string worldsheet theory, the K\"ahler parameters $\mathbf{T}$ and $\mathbf{t}$ are fugacities corresponding to fractional momenta and windings, respectively. Therefore, T-duality of this class of little string theories \cite{Intriligator:1997dh}
can be stated in the following manner
\bea
Z^{(N,M)}_{\text{IIa}}(\mathbf{T},\mathbf{t},m,\epsilon_{1,2})=Z^{(N,M)}_{\text{IIb}}(\mathbf{t},\mathbf{T},m,\epsilon_{1,2})\,,
\eea
which is indeed a symmetry of the topological partition function $\mathcal{Z}_{X_{N,M}}$ following from the double elliptic fibration structure of $X_{N,M}$.


\section{Nekrasov-Shatashvili Limit and Sigma Model}\label{Sect:SigmaModel}
In this section we study the partition function (\ref{FunctTopPartFctMN}) from the point of view of a sigma model. As we shall encounter, the NS limit suggests that there is a relation between the partition functions for generic $(N,M)$ and the particular case $(N,M)=(1,1)$.
\subsection{M-string Partition Function from the Sigma-Model Viewpoint}
Before studying the NS-limit, we first review the computation of $\mathcal{Z}_{N,M}$ defined in (\ref{FunctTopPartFctMN}) from the sigma model perspective. We will first focus on the case $M=1$ and discuss generic $M$ in section~\ref{Sect:SigmaModelGeneralisation}.

In \cite{Haghighat:2013gba} it was shown that the non-compact\footnote{In the following we mean by non-compact the limit in which the $x^6$ direction becomes non-compact, \emph{i.e.} the limit $\rho\to i\infty$.} M-strings partition function of the $(N,1)$ case can be obtained from the elliptic genus of a $(0,2)$ sigma model whose
target space is the product of Hilbert schemes of $\mathbb{C}^2$. In \cite{Hohenegger:2013ala} the general case $(N,M)$ was studied and it was shown that its partition function can be obtained from the elliptic genus of a $(0,2)$ sigma model whose target space is the product of instanton moduli spaces.

Recall that the (0,2) supersymmetric sigma model is described by a collection of scalar fields $\phi^{i}$ and $\bar{\phi}^{\bar{i}}$ on the worldsheet $\Sigma$ which are  local coordinates on the target space $M_{\text{target}}$ together with fermions which are section of various bundles on $M_{\text{target}}$:
\bea
\lambda^a \in \Gamma (\Phi^{\star} V\otimes S_+ )\,, \quad &\quad \lambda_a \in \Gamma ( \Phi^{\star} V^{\star}\otimes S_+ ) \\
\psi^i \in \Gamma (\Phi^{\star} TM \otimes S_-)\,, \quad &\quad \bar{\psi}^{\bar{i}} \in \Gamma (  \Phi^{\star} \overline{TM}\otimes S_-)\,.
\eea
Here $TM/\overline{TM}$ are holomorphic/anti-holomorphic tangent bundles, $S_{\pm}$ are the spin bundles on $\Sigma$, $\Phi:\Sigma\mapsto M_{\text{target}}$ is the map from the worldsheet to target space and $V$ is an arbitrary complex vector bundle of rank $r$ on $M_{\text{target}}$. To describe fermionic fields these bundles have first been pulled back by $\Phi^{\star}$ and tensored with spinor bundles $S_{\pm}$ of positive and negative chirality on $\Sigma$. The Lagrangian of the $(0,2)$ supersymmetric non-linear sigma model reads:
\begin{equation}
\begin{aligned}
\mathcal{L} = &\ g_{i\bar{j}} \partial_0 \phi^i \partial_0 \bar{\phi}^{\bar{j}} - g_{i\bar{j}} \partial_1 \phi^i \partial_1 \bar{\phi}^{\bar{j}} + i g_{i\bar{j}} \bar{\psi}^{\bar{j}}_- (D_0 + D_1) \psi^i + i \lambda_a (D_0 - D_1) \lambda^a \\ &+ F^a_{\ bi\bar{j}} \lambda_a \lambda^b \psi^i \bar{\psi}^{\bar{j}}
\end{aligned}
\end{equation}
where the covariant derivatives acting on $\psi^i$ and $\lambda^a$ have the following form:
\begin{equation}
\begin{aligned}
D_{\mu} \psi^i &= \partial_{\mu} \psi^i + \partial_{\mu} \phi^j \Gamma^{i}_{\ jk} \psi^k \\
D_{\mu} \lambda^a &= \partial_{\mu} \lambda^a + \partial_{\mu} \phi^i A^a_{\ bi} \lambda^b
\end{aligned}
\end{equation}
This Lagrangian has two $U(1)$ symmetries --- a left-moving flavour symmetry and a right-moving R-symmetry. We denote the conserved charge corresponding to left-moving (right-moving) $U(1)$ symmetry by $J_L (J_R)$.
Then the elliptic genus of the $(0,2)$ non-linear sigma model is defined as the following trace over the Ramond-Ramond sector of sigma model spectrum:
\begin{equation}
Z(\tau, z) = \text{Tr}_{\text{RR}} (-1)^F y^{J_L} Q_\tau^{H_L} \bar{Q}_\tau^{H_R}
\label{thetrace}
\end{equation}
where $(-1)^F = e^{i\pi (J_L - J_R)}$, $Q_\tau = e^{2\pi i \tau}$ and $y = e^{2\pi i z}$.

The relevant information in order to compute these elliptic genera is the holomorphic bundle $V$ on the respective target space $M_{\text{target}}$, to which the left-moving fermions couple. Assuming that the rank of $V$ is the same as the rank of the tangent bundle ($\text{rank}(V)=\text{rank}(TM)$), the computation of the trace in Eq.(\ref{thetrace}) using the localization principle in the path integral formalism gives the following expression for the elliptic genus:
\begin{align}
\mathcal{Z}_{M_{\text{target}}}(\tau,z)
=\int_{M_{\text{mod}}}\prod_{i=1}^{\text{rank}(TM)}\,\frac{x_i\,\vartheta(\tau,\tilde{x}_i+z)}{\vartheta(\tau,x_i)}\,,\label{DefEllGenDefInt}
\end{align}
where $\vartheta$ is defined in (\ref{DefVarthet}) and we explain below how to perform the integral for the moduli spaces mentioned above. Furthermore, $x_i$ and $\tilde{x}_i$ are the Chern-roots of the tangent bundle $TM$ and $V$ respectively, which parameterise the Chern classes as follows
\begin{align}
&c(TM)=\prod_{i=1}^{\text{rank}(TM)}(1+x_i)\,,&&\text{and} &&c(V)=\prod_{i=1}^{\text{rank}(V)}(1+\tilde{x}_i)\,.
\end{align}
The bundle $V$ was worked out explicitly in \cite{Hohenegger:2013ala} for generic $(N,M)$ and we review the salient features in the following:

To this end, we first recall some facts about the relevant target space, the instanton moduli spaces: We denote by $\mathfrak{M}_{N,k}$ the $U(N)$ instanton moduli space of charge $k$. It is well known that it has complex dimension $2Nk$  and is given by a hyperk\"ahler quotient \cite{Nakajima, Nakajima2}
\bea
\mathfrak{M}_{N,k}=\mathbb{M}_{N,k}/U(k)\,,
\eea
where
\begin{align}
\mathbb{M}_{N,k}=\{(B_{1},B_{2},i,j)~|~[B_1,B_2]  + ij =0\,,
[B_1 , B_1^{\dagger}] + [B_2, B_2^{\dagger}] - ii^{\dagger} - jj^{\dagger} = \zeta Id\}\,.
\end{align}
Here, $B_{1,2}\in \text{End}(\mathbb{C}^k)$, $i \in \text{Hom}(\mathbb{C}^k, \mathbb{C}^{N})$ and $ j \in \text{Hom}(\mathbb{C}^N, \mathbb{C}^{k})$. The $U(k)$ action is defined as
\bea g(B_1, B_2, i, j) \mapsto (g B_1  g^{-1}, g B_2 g^{-1}, g i, j g^{-1}).
\eea
Furthermore, there is a $U(1)^N \times U(1)_{\epsilon_1} \times U(1)_{\epsilon_2}$ action on $\mathfrak{M}_{N,k}$, defined as:
\begin{align}
&(B_1,B_2,i,j)\mapsto (e^{i\epsilon_1} B_{1},e^{i\epsilon_2} B_{2},i\,e^{-1}, e\,j)\,,&&\text{with} &&e=\mbox{diag}(e_1,e_2,\mathellipsis,e_N)\,.\label{action}
\end{align}
As was shown in \cite{Nakajima}, the fixed points of this action are in one-to-one correspondence with $N$-tuples of integer partitions $(\nu_1, \dots, \nu_N)$ which satisfy the condition
\begin{align}
|\nu_1| + \dots |\nu_N| = k\,,\label{CondFixedpoint}
\end{align}
where for any of the integer partitions $\nu_a=(\nu_{a,1},\ldots,\nu_{a,\ell_a})$ of length $\ell_a$ we have used the notation
\begin{align}
|\nu_a|:=\sum_{n=1}^{\ell_a}\nu_{a,n}\,.
\end{align}
The strategy advocated in \cite{Haghighat:2013gba,Hohenegger:2013ala} to compute the elliptic genus is to calculate the integral over the target space in (\ref{DefEllGenDefInt}) equivariantly, by writing it as a sum over the fixed points defined through (\ref{CondFixedpoint}). It was furthermore shown that the arguments of the $\vartheta$-functions in (\ref{DefEllGenDefInt}) can be rewritten in terms of the weights (denoted $w_{a}$ in the following) of the bundle $V$ (and $TM$ respectively) at the fixed point $(\nu_{1},\nu_{2},\ldots,\nu_{N})$ under the action (\ref{action}).

At a fixed point labelled by $(\nu_{1},\nu_{2},\ldots,\nu_{N})$, the weights $w_a$ of the above $U(1)^{N+2}$ action on the tangent bundle are
\bea\label{tangentweights}
\sum_{a}e^{w(TM)_{a}}=\sum_{a,b=1}^{N}e_{b}e_{a}^{-1} \left(\sum_{(i,j)\in \nu_{a}}q^{-(\nu_{b,j}^{t}-i)}t^{-(\nu_{a,i}-j+1)}+\sum_{(i,j)\in \nu_{b}}q^{\nu_{a,j}^{t}-i+1}t^{\nu_{b,i}-j}\right).
\label{weights1}
\eea
Here, a given partition $\nu_a$ is represented by its Young diagram and $(i,j)$ refer to the coordinates of the boxes in that diagram.

As we mentioned before the $(N,1)$ case corresponds to the target space which is the product of $N$ copies of the Hilbert scheme of points
\begin{align}
\underbrace{\mbox{Hilb}^{\bullet}[\mathbb{C}^2]\times \mbox{Hilb}^{\bullet}[\mathbb{C}^2]\times \cdots \times \mbox{Hilb}^{\bullet}[\mathbb{C}^2]}_{N-\text{copies}}\,.
\end{align}
From Eq.(\ref{weights1}), it follows that the weights of the tangent bundle in the NS-limit in this case are given by:
\bea
\lim_{t\mapsto 1}\sum_{a}e^{w(TM)_{a}}=\sum_{a=1}^{N}\left(\sum_{(i,j)\in \nu_{a}}q^{-(\nu_{a,i}-j+1)}+\sum_{(i,j)\in \nu_{a}}q^{\nu_{a,i}-j}\right)\,.
\label{NSweights1}
\eea
The bundle $V$ on the above product which couple to the left moving fermions was studied in \cite{Hohenegger:2013ala} and can be described as follows. Let $I_{\nu_{a}}$ be the fixed point corresponding to the partition $\nu_{a}$. It is a monomial ideal and the fiber of the bundle $V$ over the fixed point labelled by $(I_{\nu_{1}},I_{\nu_{2}},\cdots, I_{\nu_{N}})$ is given by \cite{Hohenegger:2013ala}
\bea
\bigoplus_{a=1}^{N}\mbox{Ext}^{1}\Big(I_{a},I_{a+1}\Big)\otimes L^{-\frac{1}{2}}\,,
\eea
where $L$ is a trivial line bundle. The weights of the bundle $V$ are given by
\bea
\sum_{a}e^{w(V)_{a}}=\sum_{a=1}^{N}Q_{m}\left(\sum_{(i,j)\in \nu_{a}}t^{\nu^{t}_{j,a+1}-i+\frac{1}{2}}\,q^{\nu_{i,a}-j+\frac{1}{2}}
+\sum_{(i,j)\in \nu_{a+1}}t^{-\nu^{t}_{j,a}+i-\frac{1}{2}}
q^{-\nu_{i,a+1}+j-
\frac{1}{2}}\right)\,.
\label{weights2}
\eea
\subsection{NS-Limit and Free Energies}
In the NS limit the weights (\ref{weights2}) become
\bea
\lim_{t\mapsto 1}\sum_{a}e^{w(V)_{a}}=\sum_{a=1}^{N}Q_{m}\left(\sum_{(i,j)\in \nu_{a}}\,q^{\nu_{i,
a}-j+\frac{1}{2}}
+\sum_{(i,j)\in \nu_{a+1}}
q^{-\nu_{i,a+1}+j-
\frac{1}{2}}\right)\,.
\label{NSweights2}
\eea
From eq.(\ref{NSweights1}) and eq.(\ref{NSweights2}), we see that the weights of the tangent bundle and the weights of $V$ are related to each other in the NS-limit:
\bea
\lim_{t\mapsto 1}\sum_{a}e^{w(V)_{a}}=\lim_{t\mapsto 1}Q_{m}\sqrt{\tfrac{q}{t}}\sum_{a}\,e^{w(TM)_{a}}\,,
\eea
\emph{i.e.} they become the same up to an over all factor. This suggests that in this limit the $(0,2)$ sigma model becomes a $(2,2)$ sigma model. Furthermore, since in this limit the target space of the $(2,2)$ model is the product of $N$ copies of the Hilbert scheme of points, the partition function, appropriately defined, becomes factorized. In fact,  it is simply given by the $(2,2)$ sigma model whose target space is the Hilbert scheme of points, which is precisely the $(1,1)$ case.

However, in the NS-limit (\ref{NSweights2}), certain weights $w(V)_a$ vanish, which leads to a divergence of the partition function, which needs to be regularised in a suitable manner. For this, several possibilities exist:
\begin{itemize}
\item One possibility, advocated in \cite{Haghighat:2015coa}, is to focus on the coefficients of particular powers of $Q_a$ in an expansion of $\mathcal{Z}_{X_{N,1}}$, which are finite in the NS-limit. The former can be extracted with the help of suitable contour integrations.
\item Another possible option is to regularise the sum over (boxes in specific) partitions $\nu_a$ in the definition of $\mathcal{Z}_{X_{N,1}}$ in such a way to exclude the vanishing weights $w(V)_a$. While in itself a very interesting possibility, it is not a priori clear that this prescription will give rise to a partition function that respects all symmetries of the brane configuration (\emph{e.g.} modularity).
\item A third possibility, advocated in our previous works \cite{Hohenegger:2015cba,Hohenegger:2015btj}, is to consider the free energy rather than the full partition function. Indeed, in the NS limit, the free energy has a pole of order $1$ in $\epsilon_2$, whose residue can be used as a definition of the free energy in the NS-limit (see eq.~(\ref{VertFreeEnergies})).
\end{itemize}
In the following we will consider the last option for the regularisation. In the partition function each factor coming from the $k$-th copy of the Hilbert scheme is weighted with the factor $Q_{k}$. However, when looking at the free energy, different contributions from the various Hilbert schemes are mixed together and are weighed with different factors. Therefore, in order to get a relation with the $(1,1)$ case we need to weigh them equally and in a uniform manner which requires taking $Q_{1}=Q_{2}=\cdots=Q_{N}$.\footnote{In section~\ref{Sect:EnhancedSym} we will discuss further reasons why this particular point of the moduli space is interesting from a physics point of view.}
\subsection{Generalisation to the General Case $(N,M)$}\label{Sect:SigmaModelGeneralisation}
The above discussion was focused on the case $M=1$. Here, we generalise the arguments to the case of generic $M$. Indeed, for an $(N,M)$ web diagram the corresponding sigma model target space is the product of instanton moduli spaces
\begin{align}
\mathfrak{M}_{M,k_{1}}\times \mathfrak{M}_{M,k_{2}}\times \cdots\times \mathfrak{M}_{M,k_{N}}\,.
\end{align}
The weights of the tangent bundle and the bundle $V$, to which left moving fermions couple, are given by
\begin{align}
\sum_{a}e^{w(TM)_{a}}&=\sum_{n=1}^{N}\sum_{a,b=1}^{M}e_{b}e_{a}^{-1}\left(
\sum_{(i,j)\in \nu^{(n)}_{a}}t^{-\nu^{t,(n)}_{b,j}+i}\,q^{-\nu^{(n)}_{a,i}+j-1}+\sum_{(i,j)\in \nu^{(n)}_{b}}t^{\nu^{t,(n)}_{a,j}-i+1}\,q^{\nu^{(n)}_{b,i}-j}\right)\,,\nonumber\\
\sum_{a}e^{w(V)_{a}}&=\sum_{n=1}^{N}\sum_{a,b=1}^{M}Q_{m}e_{b}e_{a}^{-1}\left(
\sum_{(i,j)\in \nu^{(n)}_{a}}t^{-\nu^{t,(n+1)}_{b,j}+i-\frac{1}{2}}\,q^{-\nu^{(n)}_{a,i}+j-\frac{1}{2}}+\sum_{(i,j)\in \nu^{(n+1)}_{b}}t^{\nu^{t,(n)}_{a,j}-i+\frac{1}{2}}\,q^{\nu^{(n+1)}_{b,i}-j+\frac{1}{2}}\right)
\end{align}
In the NS limit these weights simplify:
\begin{align}
\lim_{t\mapsto 1}\sum_{a}e^{w(TM)_{a}}&=\sum_{n=1}^{N}\sum_{a,b=1}^{M}e_{b}e_{a}^{-1}\left(
\sum_{(i,j)\in \nu^{(n)}_{a}}q^{-\nu^{(n)}_{a,i}+j-1}+\sum_{(i,j)\in \nu^{(n)}_{b}}q^{\nu^{(n)}_{b,i}-j}\right)\nonumber\\
\lim_{t\mapsto 1}\sum_{a}e^{w(V)_{a}}&=\sum_{n=1}^{N}\sum_{a,b=1}^{M}Q_{m}e_{b}e_{a}^{-1}\left(
\sum_{(i,j)\in \nu^{(n)}_{a}}q^{-\nu^{(n)}_{a,i}+j-\frac{1}{2}}+\sum_{(i,j)\in \nu^{(n+1)}_{b}}q^{\nu^{(n+1)}_{b,i}-j+\frac{1}{2}}\right)\,.
\end{align}
Using the cyclicity of the partitions $\nu^{(M+1)}=\nu^{(1)}$ we see from the above equation that:
\bea
\lim_{t\mapsto 1}\sum_{a}e^{w(V)_{a}}=\lim_{t\mapsto 1}Q_{m}\sqrt{\tfrac{q}{t}}\sum_{a}e^{w(T)_{a}}\,.
\eea
As in the case of $M=1$, the weights of the bundle $V$ and the weights of the tangent bundle are identical up to an overall factor. As explained above, this again suggests a relation between the free energies for generic $M,N$ and the particular case $M=N=1$ at the particular point (\ref{limit}).
\subsection{NS-limit and Point of Enhanced Symmetry}\label{Sect:EnhancedSym}
Before finishing this section and before studying the free energies explicitly in detail, we first would like to reinforce the argument that the point (\ref{limit}) together with the NS-limit is of particular physical interest.

We begin by discussing the physical significance of the NS-limit $\epsilon_2\to 0$. Recall that the free energy for any toric Calabi-Yau threefold  can be written in terms of degeneracies of BPS states, the Gopakumar-Vafa invariants, which are M2-branes wrapping the holomorphic cycles of the Calabi-Yau threefold \cite{Gopakumar:1998ii,Gopakumar:1998jq,Hollowood:2003cv},
\begin{align}
F_{X}(\omega,\epsilon_1,\epsilon_2)=-\sum_{\beta\in H_{2}(X,\mathbb{Z})}
\sum_{n=1}^{\infty}{1 \over n}
\sum_{j_{L},j_{R}}
\frac{e^{-n\,\int_{\beta}\omega}N^{j_{L},j_{R}}_{\beta}(-1)^{2j_{L}+2j_{R}}\mbox{Tr}_{j_{L}}e^{n\,j_{L,3}\epsilon_{-}}
\mbox{Tr}_{j_{R}}e^{n\,j_{R,3}\epsilon_{+}}}{(e^{\frac{n}{2}\epsilon_{1}}-e^{-\frac{n}{2}\epsilon_{1}})(e^{\frac{n}{2}\epsilon_{2}}-e^{-\frac{n}{2}\epsilon_{2}})}\,,
\end{align}
where $N_{\beta}^{j_{L},j_{R}}$ are the number of BPS states with charge $\beta\in H_{2}(X,\mathbb{Z})$ in the representation $(j_{L},j_{R})$ of $SU(2)_{L}\times SU(2)_{R}$ (the little group) and $\epsilon_{\pm}=\frac{\epsilon_{1}\pm \epsilon_{2}}{2}$. The NS limit of the free energy is given by
\bea\label{NS1}
F^{NS}_{X}=\lim_{\epsilon_{2}\mapsto 0}\epsilon_{2}F_{X}(\omega,\epsilon_1,\epsilon_2)=-\sum_{\beta\in H_{2}(X,\mathbb{Z})}
\sum_{n=1}^{\infty}{1 \over n^2}
\sum_{j}
\frac{
e^{-n\,\int_{\beta}\omega}\,n^{j}_{\beta}(-1)^{2j}
\mbox{Tr}_{j}e^{n\,j_{3}\frac{\epsilon_1}{2}}}
{(e^{n\frac{\epsilon_1}{2}}-e^{-n\frac{\epsilon_{1}}{2}})}\eea
where
\bea
n^{j}_{\beta}&=&\sum_{j_{L},j_{R}}\,
N^{j_{L},j_{R}}_{\beta}\,N^{j}_{j_{L}\,j_{R}}\,.
\eea
In the above expression $N^{j}_{j_{L}\,j_{R}}$ are the Clebsh-Gordan coefficients. Furthermore, $n^{j}_{\beta}$ is the number of particles with spin $j$ with respect to the diagonal $SU(2)\subset SU(2)_{L}\times SU(2)_{R}$ and charge $\beta$. Thus, in the NS-limit $\epsilon_2\to 0$ the coefficients $n_{\beta}^{j}$ count the number of states with charge $\beta$ and \emph{physical} spin $j$.

In the same way as the NS-limit, also the point (\ref{limit}) in moduli space is of particular interest from the point of view of counting BPS states. To understand this, we recall a number of hints, which we have found in previous work \cite{Hohenegger:2015cba,Hohenegger:2015btj} for all cases with $M=1$ (in the limit where the direction $x^6$ is non-compact), that the point
\begin{align}
t_{1}=t_{2}=\ldots=t_{N-1}\,,\label{limitM1}
\end{align}
has interesting symmetries:
\begin{itemize}
\item The NS-limit of the free energies
\begin{align}
F_{NS}^{(r_1,\ldots, r_{N-1})}(\tau,m,\epsilon_1)=\lim_{\epsilon_2\to 0}\,\epsilon_2\,\widetilde{F}^{(r_1,\ldots,r_{N-1})}(\tau,m,\epsilon_1,\epsilon_2)\,,
\end{align}
can be interpreted as counting BPS excitations of monopole-strings with magnetic charges $(r_1,\ldots,r_{N-1})$ in five dimensions. The moduli space of BPS excitations that additionally also carry electric charges, is very complicated and in general does not factorise into a center of mass and a relative contribution. In fact, as explained in \cite{Hohenegger:2015cba}, only for (\ref{limitM1}), such a factorisation is possible. Therefore, only with this specific choice, the $F_{NS}^{(r_1,\ldots, r_{N-1})}$ have an interpretation in terms of elliptic genera of the relative moduli space of magnetic monopole strings.
\item In \cite{Hohenegger:2015cba} we have seen that particular linear combinations\footnote{They will be discussed in detail in section~\ref{Sect:NonCompactFreeEnergy}.}, called $T^{(R)}$, of the (non-compact) free energies $F_{NS}^{(r_1,\ldots, r_{N-1})}(\tau,m,\epsilon_1)$ have very particular symmetry properties: Indeed, the combination $T^{(R)}$, for $R\in\mathbb{N}$, can be completely reconstructed from $T^{(1)}$ with the help of a particular combination of Hecke operators. However, such linear combinations of free energies are physically only meaningful at the particular point (\ref{limitM1}) in the moduli space.
\item Even in the limit $\epsilon_1\to 0$, the functions
\begin{align}
\mathfrak{f}_0^{(r_1,\ldots, r_{N-1})}(\tau,m)=\lim_{\epsilon_1\to 0}\,\epsilon_1\,F_{NS}^{(r_1,\ldots, r_{N-1})}(\tau,m,\epsilon_1)
\end{align}
do not quite transform as Jacobi forms under $SL(2,\mathbb{Z})$ transformations.\footnote{They are quasi-modular, \emph{i.e.} they can be made into modular forms, at the expense of not being holomorphic anymore.} However, as discussed in \cite{Hohenegger:2015cba}, for given $R\in\mathbb{N}$, the unique combination
 \begin{align}
T^{(R)}_0(\tau,m)=\sum_{\{r_a\},\sum r_a=R}\mathfrak{f}_{0}^{(r_1,\ldots,r_{N-1})}(\tau,m)\,,\label{DefTK0}
\end{align}
is a holomorphic Jacobi form of weight $-2$ and index $R$. However, again, from a physical point of view, forming combinations of different free energies as in (\ref{DefTK0}) is only meaningful for the choice (\ref{limitM1}).
\end{itemize}
We expect, upon generalising (\ref{limitM1}) to (\ref{limit}), to find similar symmetry properties also for  the (compact) free energies with $(N,M)\geq (1,1)$. As already alluded to earlier in section~\ref{Sect:BraneWebPart}, for the particular choice (\ref{limit}), the free energies become much simpler, and as anticipated in the previous sections, can in fact be fully reconstructed from the case $M=1=N$. In the following we will present computational evidence, that this is indeed the case.

\section{Relation between the Cases $(N,M)$ and $(1,1)$}\label{Sect:RelNM}
In this section we study the BPS free energies at the particular point (\ref{limit}) in the moduli space in the NS limit. We establish relations between the configurations with $M\neq 1\neq N$ and $M=1=N$.
\subsection{Horizontal Description: Case $(N,1)$ and Non-Compact Limit}\label{Sect:HorizontalHecke}
In analysing the partition function at the particular point (\ref{limit}), we first begin with the horizontal description and consider the particular case $M=1$. As we shall see, configurations with generic $N$ can be uniquely recovered from the case $M=1=N$.  Furthermore, the horizontal approach allows us to make contact to the non-compact free energies and some of the results found in \cite{Hohenegger:2015cba}. We will consider the general case $M\neq 1\neq N$ in the following section~\ref{Sect:VerticalDescription}, using the vertical description.
%
\subsubsection{Non-Compact Free Energies}\label{Sect:NonCompactFreeEnergy}
We begin by introducing some notation for the non-compact free energies, which were studied in great detail in \cite{Hohenegger:2015btj}. Specifically, we introduce the following notation for the NS-limit (for generic $N$ and $M=1$)
\begin{align}
F^{(r_1,\ldots,r_{N-1})}_{\text{NS}}(\tau,m,\epsilon_1)=\lim_{\epsilon_2\to 0}\epsilon_2\widetilde{F}^{(r_1,\ldots,r_{N-1})}(\tau,m,\epsilon_1,\epsilon_2)\,,
\end{align}
where $\widetilde{F}^{(r_1,\ldots,r_{N-1})}(\tau,m,\epsilon_1,\epsilon_2)$ are the non-compact free energies defined as in \cite{Hohenegger:2015cba}. We can further expand in $\epsilon_1$ with the coefficients defined in the following manner
\begin{align}
F^{(r_1,\ldots,r_{N-1})}_{\text{NS}}(\tau,m,\epsilon_1)=\sum_{n=0}^\infty\,\epsilon_1^{2n-1}\,\mathfrak{f}^{(r_1,\ldots,r_{N-1})}_n(\tau,m)\,.\label{NoncompactExpansionEps}
\end{align}
Explicit expressions for the first few coefficients are compiled in appendix~\ref{App:NonCompactFreeEnergies}. With the help of the free energies (\ref{NoncompactExpansionEps}) we construct the combinations (see \cite{Hohenegger:2015cba})
\begin{align}
T^{(R)}(\tau,m,\epsilon_1)=\sum_{\{r_a\},\sum r_a=R}F_{\text{NS}}^{(r_1,\ldots,r_{N-1})}(\tau,m,\epsilon_1)\,,\label{DefTK}
\end{align}
where the sum is over all configurations $(\{r_a\})=(r_1,\ldots,r_{N-1})$ with the property $\sum_{a=1}^{N-1}r_a=R$. Specifically, for the first few values of $R$
{\allowdisplaybreaks
\begin{align}
&T^{(1)}(\tau,m,\epsilon_1)=F_{\text{NS}}^{(1)}\,,\nonumber\\
&T^{(2)}(\tau,m,\epsilon_1)=F_{\text{NS}}^{(2)}+F_{\text{NS}}^{(1,1)}\,,\nonumber\\
&T^{(3)}(\tau,m,\epsilon_1)=F_{\text{NS}}^{(3)}+2\,F_{\text{NS}}^{(2,1)}+F_{\text{NS}}^{(1,1,1)}\,,\nonumber\\
&T^{(4)}(\tau,m,\epsilon_1)=F_{\text{NS}}^{(4)}+2\,F_{\text{NS}}^{(3,1)}+F_{\text{NS}}^{(2,2)}+2\,F_{\text{NS}}^{(2,1,1)}+F_{\text{NS}}^{(1,2,1)}+F_{\text{NS}}^{(1,1,1,1)}\,,
\end{align}}
where we have used the identity
\begin{align}
F_{\text{NS}}^{(r_1,\ldots,r_{N-1})}(\tau,m,\epsilon_1)=F_{\text{NS}}^{(r_{N-1},\ldots,r_{1})}(\tau,m,\epsilon_1)\,.
\end{align}
For each $T^{(R)}(\tau,m,\epsilon_1)$ we also define an expansion in powers of $\epsilon_1$
\begin{align}
T^{(R)}(\tau,m,\epsilon_1)=\sum_{n=0}^\infty\epsilon_1^{2n-1}\,\mathfrak{t}_n^{(R)}(\tau,m)\,.\label{DefExpansionTR}
\end{align}
For completeness, we have given explicit Fourier expansions of the first coefficients in appendix~\ref{App:NonCompactFreeEnergies}. The coefficient $\mathfrak{t}_0^{(R)}(\tau,m)$ is a holomorphic Jacobi form of weight $-2$ and index $R$.

\subsubsection{Connection to Non-Compact Free Energies -- Specific Examples}
The $T^{(K)}$ introduced in (\ref{DefTK}) are in fact related to the compact free energies $G^{(\{r_a\})}_{N,1}$ at the particular point (\ref{limit}) in moduli space. To understand this, we define
\begin{align}
&G^{(R)}_{N,1}(\tau,m,t_{1},\ldots,t_{N},\epsilon_1)=\lim_{\epsilon_2\to 0}\epsilon_2\,\sum_{\{r_a\},\sum r_a=R}Q_{1}^{r_1}\ldots Q_{N}^{r_N}\,G_{N,1}^{(r_1,\ldots,r_N)}(\tau,m,\epsilon_1,\epsilon_2)\,,&&\forall R>0\,,
\end{align}
and evaluate this function at the particular point (\ref{limit}) for specific values of $N$ and $R$. We begin by studying a few simple cases: For $N=1$ we have for the first few values of $R$
{\allowdisplaybreaks
\begin{align}
G_{1,1}^{(1)}&=\lim_{\epsilon_2\to 0}\epsilon_2\,Q_\rho\,G^{(1)}(\tau,m,\epsilon_1,\epsilon_2)=Q_\rho\,F_{\text{NS}}^{(1)}=Q_\rho\,T^{(1)}\,,\nonumber\\
G_{1,1}^{(2)}&=\lim_{\epsilon_2\to 0}\epsilon_2\,Q_\rho^2\,G^{(2)}(\tau,m,\epsilon_1,\epsilon_2)=Q_\rho^2\left[F_{\text{NS}}^{(2)}+F_{\text{NS}}^{(1,1)}\right]=Q_\rho^2\,T^{(2)}\,,\nonumber\\
G_{1,1}^{(3)}&=\lim_{\epsilon_2\to 0}\epsilon_2\,Q_\rho^3\,G^{(3)}(\tau,m,\epsilon_1,\epsilon_2)=Q_\rho^3\left[F_{\text{NS}}^{(3)}+2\,F_{\text{NS}}^{(2,1)}+F_{\text{NS}}^{(1,1,1)}\right]=Q_\rho^3\,T^{(3)}\,,
\end{align}}
which generalises to
\begin{align}
&G_{1,1}^{(R)}=\lim_{\epsilon_2\to 0}\epsilon_2\,Q_\rho^R\,G^{(R)}(\tau,m,\epsilon_1,\epsilon_2)=Q_\rho^R\sum_{\{r_a\},\sum r_a=R}F_{\text{NS}}^{(\{r_a\})}=Q_\rho^R\,T^{(R)}\,,&&R\in\mathbb{N}
\end{align}
where we used the conjecture in eq.~(4.17) of \cite{Hohenegger:2015cba} (supported by an extensive list of explicit checks in appendix A.2 therein) that express the compact free energies as unique linear combinations of their non-compact counterparts.

For $N=2$ we have for the first few values of $R$
{\allowdisplaybreaks
\begin{align}
G_{2,1}^{(1)}&=\lim_{\epsilon_2\to 0}\epsilon_2\,Q_\rho^{\frac{1}{2}}\,\left[G^{(1,0)}+G^{(0,1)}\right]=2\,Q_\rho^{\frac{1}{2}}\,F_{\text{NS}}^{(1)}=2\,Q_\rho^{\frac{1}{2}}\,T^{(1)}\,,\nonumber\\
G_{2,1}^{(2)}&=\lim_{\epsilon_2\to 0}\epsilon_2\,Q_\rho\,\left[G^{(2,0)}+G^{(0,2)}+G^{(1,1)}\right]=Q_\rho\,\left[2F_{\text{NS}}^{(2)}+2F_{\text{NS}}^{(1,1)}\right]=2\,Q_\rho\,T^{(2)}\,,\nonumber\\
G_{2,1}^{(3)}&=\lim_{\epsilon_2\to 0}\epsilon_2\,Q_\rho^{\frac{3}{2}}\,\left[G^{(3,0)}+G^{(0,3)}+G^{(2,1)}+G^{(1,2)}\right]=Q_\rho^{\frac{3}{2}}\,\left[2F_{\text{NS}}^{(3)}+4F_{\text{NS}}^{(2,1)}+2F_{\text{NS}}^{(1,1,1)}\right]\nonumber\\
&\hspace{1cm}=2\,Q_\rho^{\frac{3}{2}}\,T^{(3)}\,,\nonumber\\
G_{2,1}^{(4)}&=\lim_{\epsilon_2\to 0}\epsilon_2\,Q_\rho^{2}\,\left[G^{(4,0)}+G^{(0,4)}+G^{(3,1)}+G^{(1,3)}+G^{(2,2)}\right]\nonumber\\
&\hspace{1cm}=Q_\rho^{2}\,\left[2F_{\text{NS}}^{(4)}+2(2F_{\text{NS}}^{(2,1,1)}+2F_{\text{NS}}^{(3,1)})+2F^{(1,1,1,1)}_{\text{NS}}+2F^{(1,2,1)}_{\text{NS}}+2F^{(2,2)}_{\text{NS}}\right]=2\,Q_\rho^{2}\,T^{(4)}\,,
\end{align}}
For $N=3$ we have for the first few values of $R$
{\allowdisplaybreaks
\begin{align}
G_{3,1}^{(1)}&=\lim_{\epsilon_2\to 0}\epsilon_2\,Q_\rho^{\frac{1}{3}}\,\left[G^{(1,0,0)}+G^{(0,1,0)}+G^{(0,0,1)}\right]=3\,Q_\rho^{\frac{1}{3}}\,F_{\text{NS}}^{(1)}=3\,Q_\rho^{\frac{1}{3}}\,T^{(1)}\,,\nonumber\\
G_{3,1}^{(2)}&=\lim_{\epsilon_2\to 0}\epsilon_2\,Q_\rho^{\frac{2}{3}}\,\left[3G^{(2,0,0)}+3G^{(1,1,0)}\right]=3\,Q_\rho^{\frac{2}{3}}\left[F_{\text{NS}}^{(2)}+F_{\text{NS}}^{(1,1)}\right]=3\,Q_\rho^{\frac{2}{3}}\,T^{(2)}\,,\nonumber\\
G_{3,1}^{(3)}&=\lim_{\epsilon_2\to 0}\epsilon_2\,Q_\rho\,\left[3G^{(3,0,0)}+3G^{(2,1,0)}+3G^{(1,2,0)}+G^{(1,1,1)}\right]=3\,Q_\rho\left[F_{\text{NS}}^{(3)}+2F_{\text{NS}}^{(2,1)}+F_{\text{NS}}^{(1,1,1)}\right]\nonumber\\
&=3\,Q_\rho\,T^{(3)}\,.
\end{align}}
This pattern suggests the general relation\footnote{Indeed, in the next subsubsection, by using a conjectured relation \cite{Hohenegger:2015cba} between the compact and non-compact free energies, we will show that (\ref{RelHorizont1}) holds for generic $R,N\in\mathbb{N}$.}
\begin{align}
G_{N,1}^{(R)}(\tau,m,\epsilon_1)=N\,Q_\rho^{\frac{R}{N}}\,T^{(R)}(\tau,m,\epsilon_1)=N\,G_{1,1}^{(R)}(\tau,m,\epsilon_1)\,.\label{RelHorizont1}
\end{align}
This relation allows us to write the free energy (in the NS-limit) at the point (\ref{limit}) in the following manner
\begin{align}
\lim_{\epsilon_2\to 0}\epsilon_2&G_{N,1}(\tau,m,\tfrac{\rho}{N},\ldots,\tfrac{\rho}{N},\epsilon_1,\epsilon_2)=N\,W_{\emptyset\emptyset}(\tau,m,\epsilon_{1,2})+\sum_{R=1}^\infty\,G_{N,1}^{(R)}(\tau,m,\tfrac{\rho}{N},\epsilon_2)\nonumber\\
&=N\,W_{\emptyset\emptyset}(\tau,m,\epsilon_{1,2})+N\sum_{R=1}^\infty\,G_{1,1}^{(R)}(\tau,m,\tfrac{\rho}{N},\epsilon_2)\nonumber\\
&=N\lim_{\epsilon_2\to 0}\epsilon_2\,G_{1,1}(\tau,m,\tfrac{\rho}{N},\epsilon_1,\epsilon_2)\,,
\end{align}
which means that the free energy for a generic configuration $(N,1)$ can uniquely be expressed in terms of $G_{1,1}$ in the NS-limit.
 We shall refer to this property as the self-similarity of the free energies.

\subsubsection{Connection to Non-Compact Free Energies -- Generic Case $(N,1)$}
To show that the relation (\ref{RelHorizont1}) is indeed true for generic $N,R\in\mathbb{N}$, it is convenient to distinguish three separate cases: $R<N$, $R=N$ and finally $R>N.$
\begin{itemize}
\item case $R<N$\\
For generic $R,N\in\mathbb{N}$, with the restriction $R<N$ we have
{\allowdisplaybreaks\begin{align}
G^{(R)}_{N,1}&=\lim_{\epsilon_2\to 0}\,\epsilon_2\, Q_\rho^{\frac{R}{N}}\sum_{\{r_a\},\sum r_a=R}\bigg[G^{(\{r_a\},\overbrace{\text{\scriptsize{0,\ldots,0}}}^{N-\ell})}+G^{(0,\{r_a\},\overbrace{\text{\scriptsize{0,\ldots,0}}}^{N-\ell-1})}+\ldots+G^{(\overbrace{\text{\scriptsize{0,\ldots,0}}}^{N-\ell},\{r_a\})}\nonumber\\
&\hspace{2cm}+G^{(r_\ell,\overbrace{\text{\scriptsize{0,\ldots,0}}}^{N-\ell},r_1,\ldots,r_{\ell-1})}+G^{(r_{\ell-1},r_\ell,\overbrace{\text{\scriptsize{0,\ldots,0}}}^{N-\ell},r_1,\ldots,r_{\ell-2})}+\ldots+G^{(r_2,\ldots,r_{\ell},\overbrace{\text{\scriptsize{0,\ldots,0}}}^{N-\ell},r_1)}\bigg]\nonumber\\
&=N\,Q_\rho^{\frac{R}{N}}\,\sum_{\{r_a\},\sum r_a=R}F_{\text{NS}}^{(\{r_a\})}=N\,Q_\rho^{\frac{R}{N}}\,T^{(R)}\,,
\end{align}}
where the summation is over all configurations\footnote{Notice that implicitly $\ell=\ell(\{r_a\})$, \emph{i.e.} the length may be different for each $\{r_a\}$. However, in order to avoid a cluttering of the equation, we do not display the argument in the following if there is no danger of confusion.} $\{r_a\}=\{r_1,r_2,\ldots,r_\ell\}$ with $\sum_{a=1}^{\ell}=R$ such that $r_a\neq 0$ ($\forall a=1,\ldots,\ell$). Notice that, because of $R<N$, we have $\ell(\{r_a\})<N$.
\item case $R=N$\\
The previous case can be generalised to include $R=N\in\mathbb{N}$
{\allowdisplaybreaks\begin{align}
G^{(N)}_{N,1}&=\lim_{\epsilon_2\to 0}\,\epsilon_2\, Q_\rho\bigg\{\sum_{{\{r_a\},\sum r_a=N}\atop{\{r_a\}\neq (1,\ldots,1)}}\bigg[G^{(\{r_a\},\overbrace{\text{\scriptsize{0,\ldots,0}}}^{N-\ell})}+G^{(0,\{r_a\},\overbrace{\text{\scriptsize{0,\ldots,0}}}^{N-\ell-1})}+\ldots+G^{(\overbrace{\text{\scriptsize{0,\ldots,0}}}^{N-\ell},\{r_a\})}\nonumber\\
&\hspace{2cm}+G^{(r_\ell,\overbrace{\text{\scriptsize{0,\ldots,0}}}^{N-\ell},r_1,\ldots,r_{\ell-1})}+G^{(r_{\ell-1},r_\ell,\overbrace{\text{\scriptsize{0,\ldots,0}}}^{N-\ell},r_1,\ldots,r_{\ell-2})}+\ldots+G^{(r_2,\ldots,r_{\ell},\overbrace{\text{\scriptsize{0,\ldots,0}}}^{N-\ell},r_1)}\bigg]+\nonumber\\
&\hspace{2cm}+G^{(\overbrace{\text{\scriptsize{1,\ldots,1}}}^{N})}\bigg\}\nonumber\\
&=N\,Q_\rho\,\sum_{{\{r_a\},\sum r_a=N}\atop{\{r_a\}\neq (1,\ldots,1)}}F_{\text{NS}}^{(\{r_a\})}+N\,F_{\text{NS}}^{(\overbrace{\text{\scriptsize{1,\ldots,1}}}^{N})}=N\,Q_\rho\,\sum_{{\{r_a\},\sum r_a=N}}F_{\text{NS}}^{(\{r_a\})}=N\,Q_\rho\,T^{(N)}\,,
\end{align}}
where the summation in the first line excludes $(1,\ldots,1)$, such that it only contains configurations with $\ell(\{r_a\})<N$. Finally, in the last line, we have used the result of \cite{Hohenegger:2015cba}
\begin{align}
\lim_{\epsilon_2\to 0}\epsilon_2\,G^{(\overbrace{\text{\scriptsize{1,\ldots,1}}}^{N})}=N\,F_{\text{NS}}^{(\overbrace{\text{\scriptsize{1,\ldots,1}}}^{N})}\,.
\end{align}
\item case $R>N$\\
The case $R>N$ is slightly more involved and requires to distinguish different contributions
{\allowdisplaybreaks
\begin{align}
G^{(R)}_{N,1}&=\lim_{\epsilon_2\to 0}\,\epsilon_2\, Q_\rho^{\frac{R}{N}}\bigg\{\sum_{{\{r_a\},\sum r_a=R}\atop{\ell(\{r_a\})<N}}\bigg[G^{(\{r_a\},\overbrace{\text{\scriptsize{0,\ldots,0}}}^{N-\ell})}+G^{(0,\{r_a\},\overbrace{\text{\scriptsize{0,\ldots,0}}}^{N-\ell-1})}+\ldots+G^{(\overbrace{\text{\scriptsize{0,\ldots,0}}}^{N-\ell},\{r_a\})}\nonumber\\
&\hspace{2cm}+G^{(r_\ell,\overbrace{\text{\scriptsize{0,\ldots,0}}}^{N-\ell},r_1,\ldots,r_{\ell-1})}+G^{(r_{\ell-1},r_\ell,\overbrace{\text{\scriptsize{0,\ldots,0}}}^{N-\ell},r_1,\ldots,r_{\ell-2})}+\ldots+G^{(r_2,\ldots,r_{\ell},\overbrace{\text{\scriptsize{0,\ldots,0}}}^{N-\ell},r_1)}\bigg]+\nonumber\\
&\hspace{2cm}\sum_{{\{r_a\},\sum r_a=R}\atop{\ell(\{r_a\})=N}}\,G^{(\{r_a\})}\bigg\}\,.
\end{align}}
Indeed, the first term contains the contribution of all configuration of length $\ell(\{r_a\})<N$, while the second term those with length  $\ell(\{r_a\})=N$. The former contribution can be treated in the same manner as in the previous two cases, while for the latter we use the conjectured relation to the non-compact free energies in \cite{Hohenegger:2015cba}
{\allowdisplaybreaks
\begin{align}
G^{(R)}_{N,1}&=Q_\rho^{\frac{R}{N}}\bigg[N\,\sum_{{\{r_a\},\sum r_a=R}\atop{\ell(\{r_a\})<N}}\,F_{\text{NS}}^{(\{r_a\})}+\sum_{{\{r_a\},\sum r_a=R}\atop{\ell(\{r_a\})=N}}d_{(\{r_a\})}\sum_{\{m_I\},\sum m_I=R} a_{(\{m_I\})}\,F_{\text{NS}}^{(\{m_I\})}\bigg]\,,\label{RgNCaseExp}
\end{align}}
where the coefficients $a_{(\{m_I\})}$ and $d_{(\{r_a\})}$ were conjectured in \cite{Hohenegger:2015cba} to be
\begin{align}
d_{(\{r_a\})}&=\left\{\begin{array}{lcl} n = {\ell \over m} & \text{if} & (\{r_a\}) = ( \underbrace{\{r_b\}_m ,\ldots,\{r_b\}_m}_{n\text{ times}} ) \\[-18pt] \\ 1 & \  \  &\text{else} \end{array}\right.\label{Prefact}\\[10pt]
a_{(\{m_I\})}&=\left\{\begin{array}{lclc} 1 & \text{if} & r_a=\sum_{r=0}^\infty m_{I+r\ell}  \quad (a = 1, \cdots, \ell) \\ 0 & \text{else} \end{array}\right. \, . \label{Relfact}
\end{align}
Since the second term in (\ref{RgNCaseExp}) is a finite linear combination of $F_{NS}^{(\{m_I\})}$ with configurations $\{m_I\}$ of length $\ell(\{m_I\})\geq N$, we can rewrite it in the following form
\begin{align}
\sum_{{\{r_a\},\sum r_a=R}\atop{\ell(\{r_a\})=N}}d_{(\{r_a\})}\sum_{\{m_I\},\sum m_I=R} a_{(\{m_I\})}\,F_{\text{NS}}^{(\{m_I\})}=\sum_{{\{m_I\},\sum m_I=R}\atop{\ell(\{m_I\})\geq N}} c_{(\{m_I\})}\,F_{\text{NS}}^{(\{m_I\})}\,,\label{RgNrewrite}
\end{align}
for some integer coefficients $c_{(\{m_I\})}\in\mathbb{N}$, which are to be determined. To this end, for a given $\{m_I\}$ with $\sum m_I=R$ and $\ell(\{m_I\})\geq N$ we define the sequence of integers
\begin{align}
\{s_a(\{m_I\})\}=(s_1,s_2,\ldots,s_N)\,,
\end{align}
with
\begin{align}
&s_1=m_1+m_{N+1}+\ldots\,,&& s_2=m_2+m_{N+2}+\ldots\,,&&\ldots&&s_N=m_N+m_{2N}+\ldots\,.
\end{align}
Following the intuitive description of \cite{Hohenegger:2015cba}, $\{s_a(m_I)\}$ corresponds to the sequence $\{m_I\}$ 'wrapped' on a circle of circumference $N$.\footnote{For example, for $N=4$ and a configuration $m=(1,2,1,1,1,2,3)$, we have $s(m)=(2,4,4,1)$.} With this definition, for a fixed $\{m_I\}$, the coefficient $c_{(\{m_I\})}$ in (\ref{RgNrewrite}) can be written as
\begin{align}
&c_{(\{m_I\})}=\sum_{{\{r_a\},\sum r_a=R}\atop{\ell(\{r_a\})=N}}\,f_{(\{r_a\}),(\{m_I\})}\,d_{(\{r_a\})}\,,\label{Coefcrewrite}
\end{align}
where we introduced
\begin{align}
f_{(\{r_a\}),(\{m_I\})}&=\left\{\begin{array}{lclc} 1 & \text{if} & \{r_a\}\text{ is a cyclic permutation of }\{s_a(\{m_I\})\}\\ 0 & \text{else} \end{array}\right. \, .
\end{align}
Since the sum in (\ref{Coefcrewrite}) is only over distinct configurations and $d_{(\{r_a\})}$ is independent of the cyclic ordering of $\{r_a\}$, we have
\begin{align}
c_{(\{m_I\})}=d_{(\{s_a\}(\{m_I\}))}\sum_{{\{r_a\},\sum r_a=R}\atop{\ell(\{r_a\})=N}}\,f_{(\{r_a\}),(\{m_I\})}=d_{(\{s_a\}(\{m_I\}))}\,\frac{N}{d_{(\{s_a\}(\{m_I\}))}}=N\,.
\end{align}
Here we have used that for a configuration of the form (with $\{r_b\}$ a configuration of length $m\leq N$ with $\frac{N}{m}=n\in\mathbb{N}$, as in (\ref{Prefact}))
\begin{align}
\{r_a\}=( \underbrace{\{r_b\}_m ,\ldots,\{r_b\}_m}_{n\text{ times}} )\,,
\end{align}
there are only
\begin{align}
m=\frac{N}{n}=\frac{N}{d_{(\{r_a\})}}\,,
\end{align}
distinct configurations contributing in the sum in (\ref{Coefcrewrite}).
Thus, we have
\begin{align}
\sum_{{\{r_a\},\sum r_a=R}\atop{\ell(\{r_a\})=N}}d_{(\{r_a\})}\sum_{\{m_I\},\sum m_I=R} a_{(\{m_I\})}\,F_{\text{NS}}^{(\{m_I\})}=N\,\sum_{{\{m_I\},\sum m_I=R}\atop{\ell(\{m_I\})\geq N}} F_{\text{NS}}^{(\{m_I\})}\,,
\end{align}
which inserted into (\ref{RgNCaseExp}) yields
\begin{align}
G^{(R)}_{N,1}&=Q_\rho^{\frac{R}{N}}\bigg[N\,\sum_{{\{r_a\},\sum r_a=R}\atop{\ell(\{r_a\})<N}}\,F_{\text{NS}}^{(\{r_a\})}+N\,\sum_{{\{r_a\},\sum r_a=R}\atop{\ell(\{r_a\})\geq N}}\,F_{\text{NS}}^{(\{r_a\})}\bigg]=N\,Q_\rho^{\frac{R}{N}}\,\sum_{{\{r_a\},\sum r_a=R}}\,F_{\text{NS}}^{(\{r_a\})}\nonumber\\
&=N\,Q_\rho^{\frac{R}{N}}\,T^{(R)}\,.
\end{align}
\end{itemize}
Combining all three cases, we find indeed (\ref{RelHorizont1}) when using the conjectures of \cite{Hohenegger:2015cba}.

\subsubsection{Hecke Structure}\label{Sect:HeckeStructure}
\begin{figure}[h]
\begin{centering}
\vskip0.5cm
  \includegraphics[width=6in]{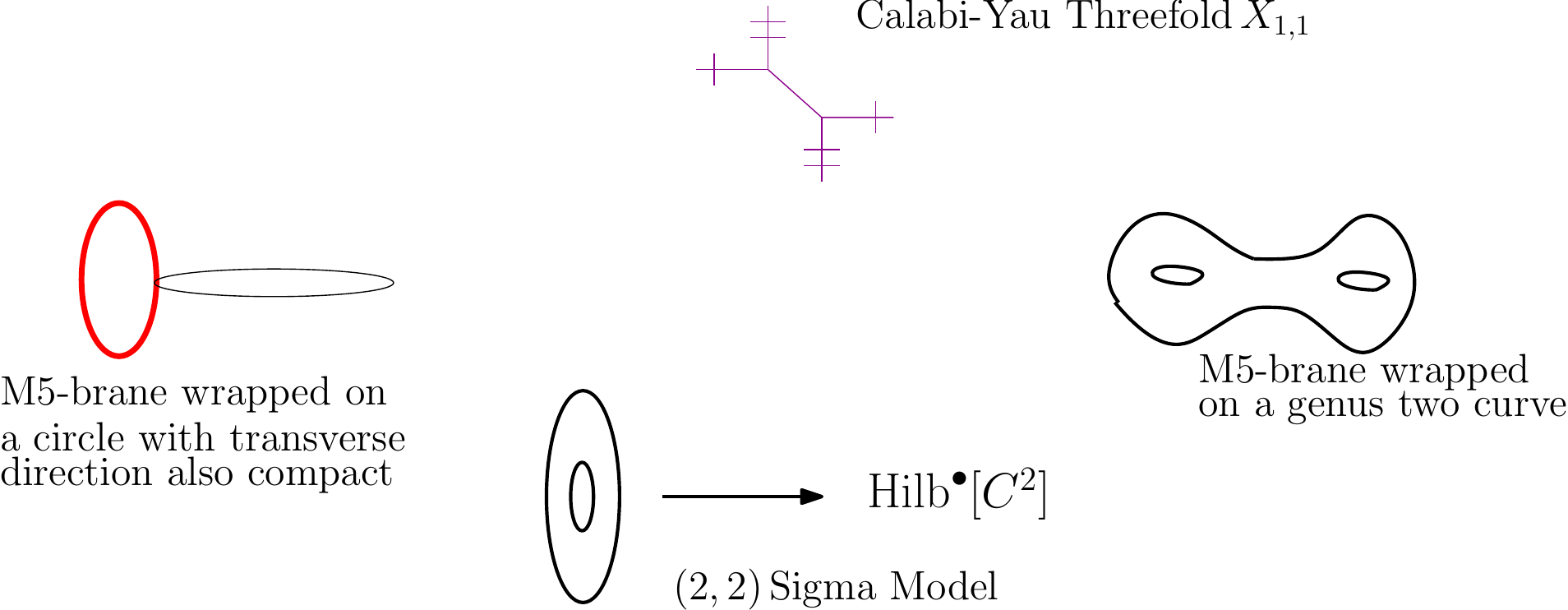}
  \vskip1cm
 \caption{\sl Various dual description of the configuration $(N,M)=(1,1)$.}\label{OneM5}
 \vskip1cm
 \end{centering}
\end{figure}

Besides (\ref{RelHorizont1}), we have in fact seen in the previous section that the free energies are related to $T^{(R)}(\tau,m,\epsilon_1)$. these quantities have already been studied in \cite{Hohenegger:2015cba}, where it was established that the $\mathfrak{t}_0^{(R)}(\tau,m)$ for $R>1$ are related to $\mathfrak{t}_0^{(1)}(\tau,m)$ through Hecke transformations, specifically
\begin{align}
\mathfrak{t}_0^{(R)}(\tau,m)=\sum_{a|R}\frac{\mu(a)}{a^3}\,\mathcal{T}_{\frac{R}{a}}\left(\mathfrak{t}_0^{(1)}(a\tau,am)\right)\,,\label{DefHEckeStructT}
\end{align}
where $\mu$ is the M\"obius function and the $R$th Hecke operator acts in the following manner on a Jacobi form $\phi(\tau,m)$ of weight $w$ and index $r$
\begin{align}
\mathcal{T}_R\left(\phi(\tau,m)\right)=R^{w-1}\sum_{{ad=R}\atop{b\text{ mod }d}}\,d^{-w}\,\phi\left(\frac{a\tau+b}{d},am\right)\,.
\end{align}
Formulated in a manner that allows comparison to (\ref{StructureProdArgument}), the result (\ref{DefHEckeStructT}) can also be stated as follows: Upon writing $\mathfrak{t}_0^{(1)}(\tau,m)$ as a Fourier series
\begin{align}
\mathfrak{t}_0^{(1)}(\tau,m)=\sum_{n=0}^\infty\sum_{\ell\in\mathbb{Z}}\,c(n,\ell)\,Q_\tau^n\,Q_m^\ell\,,
\end{align}
with coefficients $c(n,\ell)\in\mathbb{R}$, the Fourier series for $\mathfrak{t}_0^{(R)}(\tau,m)$ (with $R$ generic) is given by
\begin{align}
\mathfrak{t}_0^{(R)}(\tau,m)=\sum_{n=0}^\infty\sum_{\ell\in\mathbb{Z}}\,c(nR,\ell)\,Q_\tau^n\,Q_m^\ell\,.
\end{align}
In fact, based on explicitly checking the first few values of $K$ and $N$ (using the explicit expressions for $\mathfrak{f}_n^{(\{k_i\})}$ given in appendix~\ref{App:NonCompactFreeEnergies}), we conjecture that the same Hecke structure holds for arbitrary powers in $\epsilon_1$, \emph{i.e.} the Fourier expansion of $\mathfrak{t}_r^{(R)}(\tau,m)$ (with $r\in\mathbb{N}$) is given by
\begin{align}
\mathfrak{t}_r^{(R)}(\tau,m)=\sum_{n=0}^\infty\sum_{\ell\in\mathbb{Z}}\,c_r(nR,\ell)\,Q_\tau^n\,Q_m^\ell\,,\label{trCoefsnr}
\end{align}
with $c_0(n,\ell)=c(n,\ell)$. This follows from the fact that:\begin{align}
\Sigma^{\text{NS}}_{1,1}(\rho,\tau,m,\epsilon_{1})=\lim_{\epsilon_2\to 0}\,\epsilon_2\,\Sigma_{1,1}(\rho,\tau,m,\epsilon_{1,2})&=\sum_{R=0}^\infty Q_\rho^R\,\widetilde{G}^{(R)}_{1,1}(\tau,m,\epsilon_1)=\sum_{R=0}^\infty Q_\rho^R\,T^{(R)}(\tau,m,\epsilon_1)\,.
\end{align}
The above relation has a geometric interpretation. Recall that for $(N,M)=(1,1)$ the brane configuration is a single M5-brane wrapped on a circle with a transverse direction also compactified to a circle. The 5D theory is ${\cal N}=1$ $U(1)$ due to mass deformation. The instanton partition function of this theory is given by an equivariant index, the elliptic genus on the instanton moduli space which for this case is the Hilbert scheme of points on $\mathbb{C}^2$. This brane configuration and the corresponding theory have various dual incarnations which are described in \figref{OneM5}. \\

\noindent
The partition function in terms of the elliptic genus of the Hilbert scheme of points is given by,
\bea
{\cal Z}_{X_{1,1}}=W_{1}(\emptyset;\rho,m,\epsilon_{1,2})\sum_{k\geq 0}Q_{\tau}^{k}\,\chi_{ell}\Big(\mbox{Hilb}^{k}[\mathbb{C}^2]\Big)\,.
\eea
Where $W_{1}(\emptyset;\rho,m,\epsilon_{1,2})$ is given by (\ref{DefWNfact}). For the case of Hilbert scheme of points (or symmetric products) it is known that the above partition function can be written entirely in terms of the elliptic genus of $\mathbb{C}^2$ \cite{Dijkgraaf:1996xw,Li:2004ef},
\bea
{\cal Z}_{X_{1,1}}&=&W_{1}(\emptyset;\rho,m,\epsilon_{1,2})\sum_{k\geq 0}Q_{\tau}^{k}\,\chi_{ell}\Big(\mbox{Hilb}^{k}[\mathbb{C}^2]\Big)\,,\\\nonumber
&=&\prod_{n,k,\ell,r,s}\Big(1-Q_{\rho}^{n}Q_{\tau}^{k}Q_{m}^{\ell}q^{r}t^{s}\Big)^{C(nk,\ell,r,s)}\,,
\eea
where $C(n,\ell,r,s)$ are the Fourier coefficients in the expansion of the elliptic genus of $\mathbb{C}^2$,
\bea
\chi_{ell}(\mathbb{C}^2)=\sum_{n,\ell,r,s}C(n,\ell,r,s)Q_{\rho}^{n}Q_{m}^{\ell}q^{r}t^{s}\,.
\eea
The single particle free energy is then given by
\begin{align}
\Sigma^{\text{NS}}_{1,1}(\rho,\tau,m,\epsilon_{1})&=\lim_{\epsilon_2\to 0}\,\epsilon_2\,\text{Plog}\prod_{n,k,\ell,r,s}\Big(1-Q_{\rho}^{n}Q_{\tau}^{k}Q_{m}^{\ell}q^{r}t^{s}\Big)^{C(nk,\ell,r,s)}\,.\label{FreeEnergyEllGenC}
\end{align}
Since $C(nk,\ell,r,s)$ depends only on the product $nk$, the specific form of (\ref{FreeEnergyEllGenC}) is compatible with the structure of (\ref{trCoefsnr}).

\subsection{Vertical Description}\label{Sect:VerticalDescription}
In this section we consider the general case $M\neq 1\neq N$ using the vertical description, for which the notation is summarised in section~\ref{Sect:PartFct}. Since for the choice (\ref{limit}), the free energies only depend on one K\"ahler parameter, it is more appropriate to re-organise the Fourier expansion of the free energies (\ref{VertFreeEnergies}) in the following manner
\begin{align}
&\widetilde{\Sigma}^{(K)}_{N,M}(\tfrac{\rho}{N},\ldots,\tfrac{\rho}{N},m,\epsilon_1)=\sum_{n=0}^\infty \sum_{\ell=0}^\infty \epsilon_1^{2n-1} e^{2\pi it\,\ell}\,\mathfrak{s}_{N,M}(n,K,\ell;m)\,,&&\forall K>0\,,\label{DefTildeSigma}
\end{align}
with all the relevant information encoded in the coefficients $\mathfrak{s}_{N,M}(n,K,\ell;m)$ which have a finite Laurent series expansion in $Q_m$ with purely numerical coefficients. Examples for low
values of all the involved parameters are tabulated in appendix~\ref{Sect:CoefsVerticalExamples}.

Comparing these explicit results, we observe the remarkable pattern
\begin{align}
&\mathfrak{s}_{M,N}(n,K,\ell;m)=MN\,\mathfrak{s}_{1,1}(n,K,\ell;m)\,,\label{SymRelF11}
\end{align}
Assuming that this pattern holds for generic $M,N,n,K,\ell\in\mathbb{N}$ (with $K>0$) means that, in the limit (\ref{limit}) and the NS-limit, the BPS free energies for configurations with generic numbers of M5- and M2-branes can be fully reconstructed from the configuration with a single M5- and a single M2-brane. Indeed, inserting this pattern into (\ref{DefTildeSigma}) we have
\begin{align}
&\widetilde{\Sigma}_{N,M}^{(K)}(\tfrac{\rho}{N},\ldots,\tfrac{\rho}{N},m,\epsilon_1)=NM\,\widetilde{\Sigma}_{1,1}^{(K)}(\tfrac{\rho}{N},m,\epsilon_1)\,,&&\forall K>0\label{NMRewriteVertical}
\end{align}
which upon using (\ref{FreeEnergyNS}) leads to
\begin{align}
\lim_{\epsilon_2\to 0}\epsilon_2\,\Sigma_{N,M}(\tfrac{\rho}{N},\ldots,\tfrac{\rho}{N},\tfrac{\tau}{M},\ldots,\tfrac{\tau}{M},m,\epsilon_{1,2})&=\widetilde{\Sigma}^{(0)}_{N,M}(\tfrac{\rho}{N},m,\epsilon_1)+NM\,\sum_{K=1}^\infty e^{\frac{2\pi i\tau K}{M}}\,\widetilde{\Sigma}_{1,1}^{(K)}(\tfrac{\rho}{N},m,\epsilon_1)\nonumber\\
&=NM\,\lim_{\epsilon_2\to 0}\epsilon_2\,\Sigma_{1,1}(\tfrac{\rho}{N},\tfrac{\tau}{M},m,\epsilon_{1,2})\,.\label{ReplicationVertical}
\end{align}
Here, for the term $\widetilde{\Sigma}^{(0)}_{N,M}$ we have used the definition (\ref{DefWNfact}) together with
\begin{align}
&Q_{a,a+b}\big|_{t_1=t_2=\ldots=t_N}=Q_\rho^{\frac{b}{N}}\,,&&\forall 0\leq a,b\leq N\,,
\end{align}
which is independent of $a$, such that at the point (\ref{limit})
\begin{align}
(W_N(\emptyset;&\tfrac{\rho}{N},\tfrac{\rho}{N},\ldots,\tfrac{\rho}{N},m,\epsilon_{1,2}))^M\nonumber\\
&=f(\rho)\,\left[\prod_{i,j,k}\prod_{b=1}^N\frac{1-Q_\rho^{\frac{N(k-1)+b}{N}}Q_m^{-1}t^{i-\frac{1}{2}}q^{j-\frac{1}{2}}}{1-Q_\rho^{\frac{N(k-1)+b}{N}}\,t^{i-1}q^{j}}\,\frac{1-Q_\rho^{\frac{N(k-1)+b-1}{N}}Q_m\, t^{i-\frac{1}{2}}q^{j-\frac{1}{2}}}{1-Q_\rho^{\frac{N(k-1)+b}{N}}\,t^{i}q^{j-1}}\right]^{MN}\,.
\end{align}
Here $f(\rho)$ is a term independent of $\epsilon_2$ which does not contribute in the NS-limit of the free energy, such that
\begin{align}
\widetilde{\Sigma}^{(0)}_{N,M}(\tfrac{\rho}{N},m,\epsilon_1)=NM\,\widetilde{\Sigma}^{(0)}_{1,1}(\tfrac{\rho}{N},m,\epsilon_1)\,.
\end{align}
Relation (\ref{ReplicationVertical}) therefore is indeed a direct relation between the free energies for generic $(N,M)$ and $(1,1)$. This property generalizes the self-similarity we already observed to arbitrary $M>1$.

From the point of view of little string theory, relation (\ref{ReplicationVertical}) implies that the partition function is self-dual under T-duality for the particular choice (\ref{limit}). Indeed, since
\begin{align}
&\mathfrak{s}_{1,1}(n,K,\ell;m)=\mathfrak{s}_{1,1}(n,\ell,K;m)\,,&&\forall n,K,\ell\in\mathbb{N}\,\text{ with }K\ell\neq 0\,,\label{ExchangeSymmetry}
\end{align}
due to the exchange symmetry of $\rho$ and $\tau$ for $M=N$, (\ref{SymRelF11}) implies that
\begin{align}
&\mathfrak{s}_{M,N}(n,K,\ell;m)=\mathfrak{s}_{M,N}(n,\ell,K;m)\,,&&\forall n,K,\ell\in\mathbb{N}\,\text{ with }K\ell\neq 0\,,
\end{align}
even for $M\neq N$. In fact, (\ref{ExchangeSymmetry}) (which is a consequence of fiber-base duality from the geometric perspective), is a particular realisation of  a more general pattern emerging in the examples of appendix~\ref{FreeEnergiesM1N1}. Indeed, for $K_1\ell_1\neq 0\neq K_2\ell_2$ we observe
\begin{align}
&\mathfrak{s}_{1,1}(n,K_1,\ell_1;m)=\mathfrak{s}_{1,1}(n,K_2,\ell_2;m)&&\Longleftrightarrow && K_1\ell_1=K_2\ell_2\,,&&\forall n\in\mathbb{Z}\,,\label{StructureProdArgument}
\end{align}
which through (\ref{SymRelF11}) also appears for generic $(N,M)$. As discussed in section~\ref{Sect:HeckeStructure}, this generalises the Hecke structure  found in \cite{Hohenegger:2015btj} for non-compact free energies (\emph{i.e.} for which the $x^6$-direction is decompactified).

\section{Conclusions}\label{Sect:Conclusions}
In this paper, we studied a class of little string theories with $8$ supercharges which can be realized through systems of $M$ parallel M5-branes probing a transverse $A_{N-1}$ singularity or $N$ M5-branes probing an $A_{M-1}$ singularity. These two descriptions are dual to each other. Furthermore, through a chain of dualities, the above brane configurations can be related to a class of toric, non-compact Calabi-Yau threefolds $X_{N,M}$. The little string partition functions are then captured by the topological partition function $\mathcal{Z}_{X_{N,M}}$ of $X_{N,M}$, which was computed in \cite{Hohenegger:2013ala} for generic $(N,M)$ using different methods. Using the M-string worldsheet description, we observe that $\mathcal{Z}_{X_{N,M}}$ potentially factorises in the NS-limit, however, at the same time becomes divergent. In order to regularise these divergences, we considered the free energies $\Sigma_{N,M}(\mathbf{t},\mathbf{T},m,\epsilon_{1,2})$ in the particular point (\ref{limit}), which we motivated by different physical considerations \cite{Hohenegger:2015cba}. Through explicit computations, we observe that the NS-limit of $\Sigma_{N,M}(\tfrac{\rho}{N},\ldots,\tfrac{\rho}{N},\tfrac{\tau}{M},\ldots,\tfrac{\tau}{M},m,\epsilon_{1,2})$ for generic $N,M\in\mathbb{N}$ is identical (up to a prefactor $NM$) to the NS-limit of $\Sigma_{1,1}(\tfrac{\rho}{N},\tfrac{\tau}{M},m,\epsilon_{1,2})$ (see (\ref{ReplicationVertical})). From a physical point of view, this means that the degeneracies of the BPS states for generic $(N,M)$ can be uniquely be reconstructed from the configuration $(N,M)=(1,1)$, a property which we termed \emph{self-similarity}. Moreover, since $\Sigma_{1,1}(\tfrac{\rho}{N},\tfrac{\tau}{M},m,\epsilon_{1,2})=\Sigma_{1,1}(\tfrac{\tau}{M},\tfrac{\rho}{N},m,\epsilon_{1,2})$, relation (\ref{ReplicationVertical}) implies that the orbifolds of the little string theories are self-dual under T-duality at (\ref{limit}). Finally, we observe a generalisation of a recursive structure, which was first established in \cite{Hohenegger:2015cba} for the free energies of non-compact M-string configurations. This structure is explained by the fact that $\Sigma_{1,1}(\tfrac{\rho}{N},\tfrac{\tau}{M},m,\epsilon_{1,2})$ is related to the (regularised) elliptic genus of $\mathbb{C}^2$.

In this work we have focused on $\mathbb{Z}_M$ orbifolds of little string theories of $A_{N-1}$ type with 8 supercharges. It would be very interesting to generalise this discussion to orbifolding little string theories of type $G$ by a generic group $H$. One question is to understand whether there are any regions in the moduli space of such theories in which the partition function displays similar symmetry properties as we discussed in this paper. Finally, it would also be very interesting to analyse symmetries of the partition functions of little string theories with $\mathcal{N}=(1,0)$ symmetry that correspond to M5-branes probing a transverse space that is not an orbifold (but is different from $\mathbb{R}^4$).

\section*{Acknowledgement}
We thank C.~Vafa for helpful discussions. SJR was supported in part by the National Research Foundation of Korea grants 2005-0093843, 2010-220-C00003 and 2012K2A1A9055280. The work of SH was partly supported by the BQR Accueil EC 2015. A.I. was supported in part by the Higher Education Commission grant HEC-20-2518.

\appendix
%

\section{Free Energies}\label{App:FreeEnergies}
In this appendix we provide explicit expressions for the expansion coefficients of the free energies $\widetilde{\Sigma}^{(K)}_{N,M}$ and $\widetilde{G}^{(R)}_{N,M}$. These will act as very non-trivial checks for the conjecture made in section~\ref{Sect:RelNM} that at the particular point (\ref{limit}), the free energies for $(N,M)\geq (1,1)$ can be uniquely constructed from the case $M=1=N$.
\subsection{Vertical Description}\label{Sect:CoefsVerticalExamples}
We first consider the expansion coefficients $\mathfrak{s}_{N,M}(n,K,\ell;m)$ in the vertical description, as defined in (\ref{DefTildeSigma}). To save writing, we will suppress the dependence on the remaining parameter $Q_m$ in the following. With this notation, we can tabulate a few of the $\mathfrak{s}_{N,M}(K,n,\ell)$ for different values of $M$ and $N$.
\subsubsection{Expansion Coefficients for $M=1$ and $N=1$}\label{FreeEnergiesM1N1}
As a reference, we first provide the coefficients for the case $M=1=N$. The coefficients of later computations will be compared to these expressions.

For $n=0$ we have for the first few coefficients
{\allowdisplaybreaks\begin{align}
\mathfrak{s}_{1,1}(0,1,0)&=2-\frac{1}{Q_m}-Q_m\,,\nonumber\\
\mathfrak{s}_{1,1}(0,1,1)&=2 Q_m ^2+\frac{2}{Q_m ^2}-8 Q_m -\frac{8}{Q_m }+12\,,\nonumber\\
\mathfrak{s}_{1,1}(0,1,2)&=-Q_m ^3-\frac{1}{Q_m ^3}+12 Q_m ^2+\frac{12}{Q_m ^2}-39 Q_m -\frac{39}{Q_m }+56\,,\nonumber\\
\mathfrak{s}_{1,1}(0,1,3)&=-8 Q_m ^3-\frac{8}{Q_m ^3}+56 Q_m ^2+\frac{56}{Q_m ^2}-152 Q_m -\frac{152}{Q_m }+208\,,\nonumber\\
\mathfrak{s}_{1,1}(0,1,4)&=2 Q_m ^4+\frac{2}{Q_m ^4}-39 Q_m ^3-\frac{39}{Q_m ^3}+208 Q_m ^2+\frac{208}{Q_m ^2}-513
   Q_m -\frac{513}{Q_m }+684\,,\nonumber\\
\mathfrak{s}_{1,1}(0,1,5)&=12 Q_m ^4+\frac{12}{Q_m ^4}-152 Q_m ^3-\frac{152}{Q_m ^3}+684 Q_m ^2+\frac{684}{Q_m ^2}-1560
   Q_m -\frac{1560}{Q_m }+2032\,,\nonumber\\
\mathfrak{s}_{1,1}(0,1,6)&=-Q_m ^5-\frac{1}{Q_m ^5}+56 Q_m ^4+\frac{56}{Q_m ^4}-513 Q_m ^3-\frac{513}{Q_m ^3}+2032
   Q_m ^2+\frac{2032}{Q_m ^2}-4382 Q_m \nonumber\\
   &\hspace{0.5cm}-\frac{4382}{Q_m }+5616\,,\nonumber\\
\mathfrak{s}_{1,1}(0,1,7)&=-8 Q_m ^5-\frac{8}{Q_m ^5}+208 Q_m ^4+\frac{208}{Q_m ^4}-1560 Q_m ^3-\frac{1560}{Q_m ^3}+5616
   Q_m ^2+\frac{5616}{Q_m ^2}-11552 Q_m\nonumber\\
   &\hspace{0.5cm} -\frac{11552}{Q_m }+14592\,,\nonumber\\
\mathfrak{s}_{1,1}(0,1,8)&=-39 Q_m ^5-\frac{39}{Q_m ^5}+684 Q_m ^4+\frac{684}{Q_m ^4}-4382 Q_m ^3-\frac{4382}{Q_m ^3}+14592
   Q_m ^2+\frac{14592}{Q_m ^2}-28899 Q_m\nonumber\\
   &\hspace{0.5cm} -\frac{28899}{Q_m }+36088\,,\nonumber\\
\mathfrak{s}_{1,1}(0,1,9)&=2 Q_m ^6+\frac{2}{Q_m ^6}-152 Q_m ^5-\frac{152}{Q_m ^5}+2032 Q_m ^4+\frac{2032}{Q_m ^4}-11552
   Q_m ^3-\frac{11552}{Q_m ^3}\nonumber\\
   &\hspace{0.5cm}+36088 Q_m ^2+\frac{36088}{Q_m ^2}-69168 Q_m -\frac{69168}{Q_m }+85500\,,\nonumber\\[10pt]
\mathfrak{s}_{1,1}(0,2,0)&=2-\frac{1}{Q_m}-Q_m\,,\nonumber\\
\mathfrak{s}_{1,1}(0,2,1)&=-Q_m ^3-\frac{1}{Q_m ^3}+12 Q_m ^2+\frac{12}{Q_m ^2}-39 Q_m -\frac{39}{Q_m }+56\,,\nonumber\\
\mathfrak{s}_{1,1}(0,2,2)&=2 Q_m ^4+\frac{2}{Q_m ^4}-39 Q_m ^3-\frac{39}{Q_m ^3}+208 Q_m ^2+\frac{208}{Q_m ^2}-513
   Q_m -\frac{513}{Q_m }+684\,,\nonumber\\
\mathfrak{s}_{1,1}(0,2,3)&=-Q_m ^5-\frac{1}{Q_m ^5}+56 Q_m ^4+\frac{56}{Q_m ^4}-513 Q_m ^3-\frac{513}{Q_m ^3}+2032
   Q_m ^2+\frac{2032}{Q_m ^2}-4382 Q_m \nonumber\\
   &\hspace{0.5cm}-\frac{4382}{Q_m }+5616\,,\nonumber\\
\mathfrak{s}_{1,1}(0,2,4)&=-39 Q_m ^5-\frac{39}{Q_m ^5}+684 Q_m ^4+\frac{684}{Q_m ^4}-4382 Q_m ^3-\frac{4382}{Q_m ^3}+14592
   Q_m ^2+\frac{14592}{Q_m ^2}\nonumber\\
&\hspace{0.5cm}-28899 Q_m -\frac{28899}{Q_m }+36088\,,\nonumber\\[10pt]
\mathfrak{s}_{1,1}(1,3,0)&=2-\frac{1}{Q_m}-Q_m\,,\nonumber\\
\mathfrak{s}_{1,1}(1,3,1)&=-8 Q_m ^3-\frac{8}{Q_m ^3}+56 Q_m ^2+\frac{56}{Q_m ^2}-152 Q_m -\frac{152}{Q_m }+208\,,\nonumber\\
\mathfrak{s}_{1,1}(1,3,2)&=-Q_m ^5-\frac{1}{Q_m ^5}+56 Q_m ^4+\frac{56}{Q_m ^4}-513 Q_m ^3-\frac{513}{Q_m ^3}+2032Q_m ^2+\frac{2032}{Q_m ^2}-4382 Q_m \nonumber\\
&\hspace{0.5cm}-\frac{4382}{Q_m }+5616\,,\nonumber\\
\mathfrak{s}_{1,1}(1,3,3)&=2 Q_m ^6+\frac{2}{Q_m ^6}-152 Q_m ^5-\frac{152}{Q_m ^5}+2032 Q_m ^4+\frac{2032}{Q_m ^4}-11552
   Q_m ^3-\frac{11552}{Q_m ^3}\nonumber\\
 &\hspace{0.5cm}+36088 Q_m ^2+\frac{36088}{Q_m ^2}-69168 Q_m -\frac{69168}{Q_m }+85500\,,\nonumber\\
\mathfrak{s}_{1,1}(1,3,4)&=-Q_m ^7-\frac{1}{Q_m ^7}+208 Q_m ^6+\frac{208}{Q_m ^6}-4382 Q_m ^5-\frac{4382}{Q_m ^5}+36088
   Q_m ^4+\frac{36088}{Q_m ^4}\nonumber\\
   &\hspace{0.5cm}-159372 Q_m ^3-\frac{159372}{Q_m ^3}+431984
   Q_m ^2+\frac{431984}{Q_m ^2}-768885 Q_m -\frac{768885}{Q_m }+928720\,.
\end{align}}
For $n=1$ we have explicitly
{\allowdisplaybreaks\begin{align}
\mathfrak{s}_{1,1}(1,1,0)&=\frac{1}{24}\left(Q_m +\frac{1}{Q_m }+4\right)\,,\nonumber\\
\mathfrak{s}_{1,1}(1,1,1)&=\frac{1}{24}\left(4 Q_m ^2+\frac{4}{Q_m ^2}-40 Q_m -\frac{40}{Q_m }+72\right)\,,\nonumber\\
\mathfrak{s}_{1,1}(1,1,2)&=\frac{1}{24}\left(Q_m ^3+\frac{1}{Q_m ^3}+72 Q_m ^2+\frac{72}{Q_m ^2}-369 Q_m -\frac{369}{Q_m }+592\right)\,,\nonumber\\
\mathfrak{s}_{1,1}(1,1,3)&=\frac{1}{24}\left(-40 Q_m ^3-\frac{40}{Q_m ^3}+592 Q_m ^2+\frac{592}{Q_m ^2}-2104 Q_m -\frac{2104}{Q_m }+3104\right)\,,\nonumber\\
\mathfrak{s}_{1,1}(1,1,4)&=\frac{1}{24}\left(4 Q_m ^4+\frac{4}{Q_m ^4}-369 Q_m ^3-\frac{369}{Q_m ^3}+3104 Q_m ^2+\frac{3104}{Q_m ^2}-9327
   Q_m -\frac{9327}{Q_m }+13176\right)\,,\nonumber\\
\mathfrak{s}_{1,1}(1,1,5)&=\frac{1}{24}\bigg(72 Q_m ^4+\frac{72}{Q_m ^4}-2104 Q_m ^3-\frac{2104}{Q_m ^3}+13176 Q_m ^2+\frac{13176}{Q_m ^2}-35064
   Q_m -\frac{35064}{Q_m }+47840\bigg)\,,\nonumber\\
\mathfrak{s}_{1,1}(1,1,6)&=\frac{1}{24}\bigg(Q_m ^5+\frac{1}{Q_m ^5}+592 Q_m ^4+\frac{592}{Q_m ^4}-9327 Q_m ^3-\frac{9327}{Q_m ^3}+47840
   Q_m ^2+\frac{47840}{Q_m ^2}-117202 Q_m \nonumber\\
   &\hspace{0.5cm}-\frac{117202}{Q_m }+156192\bigg)\,,\nonumber\\
\mathfrak{s}_{1,1}(1,1,7)&=\frac{1}{24}\bigg(-40 Q_m ^5-\frac{40}{Q_m ^5}+3104 Q_m ^4+\frac{3104}{Q_m ^4}-35064 Q_m ^3-\frac{35064}{Q_m ^3}+156192
   Q_m ^2+\frac{156192}{Q_m ^2}\nonumber\\
   &\hspace{0.5cm}-358048 Q_m -\frac{358048}{Q_m }+467712\bigg)\,,\nonumber\\
\mathfrak{s}_{1,1}(1,1,8)&=\frac{1}{24}\bigg(-369 Q_m ^5-\frac{369}{Q_m ^5}+13176 Q_m ^4+\frac{13176}{Q_m ^4}-117202
   Q_m ^3-\frac{117202}{Q_m ^3}+467712 Q_m ^2\nonumber\\
   &\hspace{0.5cm}+\frac{467712}{Q_m ^2}-1017981
   Q_m -\frac{1017981}{Q_m }+1309328\bigg)\,,\nonumber\\
\mathfrak{s}_{1,1}(1,1,9)&=\frac{1}{24}\bigg(4 Q_m ^6+\frac{4}{Q_m ^6}-2104 Q_m ^5-\frac{2104}{Q_m ^5}+47840 Q_m ^4+\frac{47840}{Q_m ^4}-358048
   Q_m ^3-\frac{358048}{Q_m ^3}\nonumber\\
   &\hspace{0.5cm}+1309328 Q_m ^2+\frac{1309328}{Q_m ^2}-2728176
   Q_m -\frac{2728176}{Q_m }+3462312\bigg)\,,\nonumber\\[10pt]
\mathfrak{s}_{1,1}(1,2,0)&=\frac{1}{24}\left(Q_m +\frac{1}{Q_m }+4\right)\,,\nonumber\\
\mathfrak{s}_{1,1}(1,2,1)&=\frac{1}{24}\left(Q_m ^3+\frac{1}{Q_m ^3}+72 Q_m ^2+\frac{72}{Q_m ^2}-369 Q_m -\frac{369}{Q_m }+592\right)\,,\nonumber\\
\mathfrak{s}_{1,1}(1,2,2)&=\frac{1}{24}\left(4 Q_m ^4+\frac{4}{Q_m ^4}-369 Q_m ^3-\frac{369}{Q_m ^3}+3104 Q_m ^2+\frac{3104}{Q_m ^2}-9327
   Q_m -\frac{9327}{Q_m }+13176\right)\,,\nonumber\\
\mathfrak{s}_{1,1}(1,2,3)&=\frac{1}{24}\bigg(Q_m ^5+\frac{1}{Q_m ^5}+592 Q_m ^4+\frac{592}{Q_m ^4}-9327 Q_m ^3-\frac{9327}{Q_m ^3}+47840
   Q_m ^2+\frac{47840}{Q_m ^2}\nonumber\\
&\hspace{0.5cm}-117202 Q_m -\frac{117202}{Q_m }+156192\bigg)\,,\nonumber\\
\mathfrak{s}_{1,1}(1,2,4)&=\frac{1}{24}\bigg(-369 Q_m ^5-\frac{369}{Q_m ^5}+13176 Q_m ^4+\frac{13176}{Q_m ^4}-117202
   Q_m ^3-\frac{117202}{Q_m ^3}+467712 Q_m ^2\nonumber\\
   &\hspace{0.5cm}+\frac{467712}{Q_m ^2}-1017981
   Q_m -\frac{1017981}{Q_m }+1309328\bigg)\,,\nonumber\\[10pt]
\mathfrak{s}_{1,1}(1,3,0)&=\frac{1}{24}\left(Q_m +\frac{1}{Q_m }+4\right)\,,\nonumber\\
\mathfrak{s}_{1,1}(1,3,1)&=\frac{1}{24}\left(-40 Q_m ^3-\frac{40}{Q_m ^3}+592 Q_m ^2+\frac{592}{Q_m ^2}-2104 Q_m -\frac{2104}{Q_m }+3104\right)\,,\nonumber\\
\mathfrak{s}_{1,1}(1,3,2)&=\frac{1}{24}\bigg(Q_m ^5+\frac{1}{Q_m ^5}+592 Q_m ^4+\frac{592}{Q_m ^4}-9327 Q_m ^3-\frac{9327}{Q_m ^3}+47840
   Q_m ^2+\frac{47840}{Q_m ^2}\nonumber\\
   &\hspace{0.5cm}-117202 Q_m -\frac{117202}{Q_m }+156192\bigg)\,,\nonumber\\
\mathfrak{s}_{1,1}(1,3,3)&=\frac{1}{24}\bigg(4 Q_m ^6+\frac{4}{Q_m ^6}-2104 Q_m ^5-\frac{2104}{Q_m ^5}+47840 Q_m ^4+\frac{47840}{Q_m ^4}-358048
   Q_m ^3-\frac{358048}{Q_m ^3}\nonumber\\
&\hspace{0.5cm}+1309328 Q_m ^2+\frac{1309328}{Q_m ^2}-2728176
   Q_m -\frac{2728176}{Q_m }+3462312\bigg)\,,
\end{align}}
\subsubsection{Expansion Coefficients for $M=2$ and $N=1$}
For $n=0$ we have for the first few coefficients
{\allowdisplaybreaks\begin{align}
\mathfrak{s}_{1,2}(0,1,0)&=2\left(2-\frac{1}{Q_m}-Q_m\right)\,,\nonumber\\
\mathfrak{s}_{1,2}(0,1,1)&=2\left(2 Q_m ^2+\frac{2}{Q_m ^2}-8 Q_m -\frac{8}{Q_m }+12\right)\,,\nonumber\\
\mathfrak{s}_{1,2}(0,1,2)&=2\left(-Q_m ^3-\frac{1}{Q_m ^3}+12 Q_m ^2+\frac{12}{Q_m ^2}-39 Q_m -\frac{39}{Q_m }+56\right)\,,\nonumber\\
\mathfrak{s}_{1,2}(0,2,0)&=2\left(2-\frac{1}{Q_m}-Q_m\right)\,,\nonumber\\
\mathfrak{s}_{1,2}(0,2,1)&=2\left(-Q_m ^3-\frac{1}{Q_m ^3}+12 Q_m ^2+\frac{12}{Q_m ^2}-39 Q_m -\frac{39}{Q_m }+56\right)\,,\nonumber\\
\mathfrak{s}_{1,2}(0,2,2)&=2\left(2 Q_m ^4+\frac{2}{Q_m ^4}-39 Q_m ^3-\frac{39}{Q_m ^3}+208 Q_m ^2+\frac{208}{Q_m ^2}-513
   Q_m -\frac{513}{Q_m }+684\right)\,.
\end{align}}
For the next-to-leading term in $\epsilon_1$ (\emph{i.e.} $n=1$) we have
{\allowdisplaybreaks \begin{align}
\mathfrak{s}_{1,2}(1,1,0)&=\frac{1}{12}\left(Q_m +\frac{1}{Q_m }+4\right)\,,\nonumber\\
\mathfrak{s}_{1,2}(1,1,1)&=\frac{1}{12}\left(4 Q_m ^2+\frac{4}{Q_m ^2}-40 Q_m -\frac{40}{Q_m }+72\right)\,,\nonumber\\
\mathfrak{s}_{1,2}(1,1,2)&=\frac{1}{12}\left(Q_m ^3+\frac{1}{Q_m ^3}+72 Q_m ^2+\frac{72}{Q_m ^2}-369 Q_m -\frac{369}{Q_m }+592\right)\,,\nonumber\\
\mathfrak{s}_{1,2}(1,2,0)&=\frac{1}{12}\left(Q_m +\frac{1}{Q_m }+4\right)\,,\nonumber\\
\mathfrak{s}_{1,2}(1,2,1)&=\frac{1}{12}\left(Q_m ^3+\frac{1}{Q_m ^3}+72 Q_m ^2+\frac{72}{Q_m ^2}-369 Q_m -\frac{369}{Q_m }+592\right)\,,\nonumber\\
\mathfrak{s}_{1,2}(1,2,2)&=\frac{1}{12}\left(4 Q_m ^4+\frac{4}{Q_m ^4}-369 Q_m ^3-\frac{369}{Q_m ^3}+3104 Q_m ^2+\frac{3104}{Q_m ^2}-9327
   Q_m -\frac{9327}{Q_m }+13176\right)\,.
\end{align}}
\subsubsection{Expansion Coefficients for $M=2$ and $N=2$}
For $n=0$ we have for the first few coefficients
{\allowdisplaybreaks\begin{align}
\mathfrak{s}_{2,2}(0,1,0)&=4\left(2-\frac{1}{Q_m}-Q_m\right)\,,\nonumber\\
\mathfrak{s}_{2,2}(0,1,1)&=4\left(2 Q_m ^2+\frac{2}{Q_m ^2}-8 Q_m -\frac{8}{Q_m }+12\right)\,,\nonumber\\
\mathfrak{s}_{2,2}(0,1,2)&=4\left(-Q_m ^3-\frac{1}{Q_m ^3}+12 Q_m ^2+\frac{12}{Q_m ^2}-39 Q_m -\frac{39}{Q_m }+56\right)\,,\nonumber\\
\mathfrak{s}_{2,2}(0,2,0)&=4\left(2-\frac{1}{Q_m}-Q_m\right)\,,\nonumber\\
\mathfrak{s}_{2,2}(0,2,1)&=4\left(-Q_m ^3-\frac{1}{Q_m ^3}+12 Q_m ^2+\frac{12}{Q_m ^2}-39 Q_m -\frac{39}{Q_m }+56\right)\,,\nonumber\\
\mathfrak{s}_{2,2}(0,2,2)&=4\left(2 Q_m ^4+\frac{2}{Q_m ^4}-39 Q_m ^3-\frac{39}{Q_m ^3}+208 Q_m ^2+\frac{208}{Q_m ^2}-513
   Q_m -\frac{513}{Q_m }+684\right)\,.
\end{align}}
For $n=1$ we have explicitly
{\allowdisplaybreaks\begin{align}
\mathfrak{s}_{2,2}(1,1,0)&=\frac{1}{6}\left(Q_m +\frac{1}{Q_m }+4\right)\,,\nonumber\\
\mathfrak{s}_{2,2}(1,1,1)&=\frac{1}{6}\left(4 Q_m ^2+\frac{4}{Q_m ^2}-40 Q_m -\frac{40}{Q_m }+72\right)\,,\nonumber\\
\mathfrak{s}_{2,2}(1,1,2)&=\frac{1}{6}\left(Q_m ^3+\frac{1}{Q_m ^3}+72 Q_m ^2+\frac{72}{Q_m ^2}-369 Q_m -\frac{369}{Q_m }+592\right)\,,\nonumber\\
\mathfrak{s}_{2,2}(1,2,0)&=\frac{1}{6}\left(Q_m +\frac{1}{Q_m }+4\right)\,,\nonumber\\
\mathfrak{s}_{2,2}(1,2,1)&=\frac{1}{6}\left(Q_m ^3+\frac{1}{Q_m ^3}+72 Q_m ^2+\frac{72}{Q_m ^2}-369 Q_m -\frac{369}{Q_m }+592\right)\,,\nonumber\\
\mathfrak{s}_{2,2}(1,2,2)&=\frac{1}{6}\left(4 Q_m ^4+\frac{4}{Q_m ^4}-369 Q_m ^3-\frac{369}{Q_m ^3}+3104 Q_m ^2+\frac{3104}{Q_m ^2}-9327
   Q_m -\frac{9327}{Q_m }+13176\right)\,.
\end{align}}
\subsubsection{Expansion Coefficients for $M=2$ and $N=3$}
For $n=0$ we have for the first few coefficients
{\allowdisplaybreaks\begin{align}
\mathfrak{s}_{3,2}(0,1,0)&=6\left(2-\frac{1}{Q_m}-Q_m\right)\,,\nonumber\\
\mathfrak{s}_{3,2}(0,1,1)&=6\left(2 Q_m ^2+\frac{2}{Q_m ^2}-8 Q_m -\frac{8}{Q_m }+12\right)\,,\nonumber\\
\mathfrak{s}_{3,2}(0,1,2)&=6\left(-Q_m ^3-\frac{1}{Q_m ^3}+12 Q_m ^2+\frac{12}{Q_m ^2}-39 Q_m -\frac{39}{Q_m }+56\right)\,,\nonumber\\
\mathfrak{s}_{3,2}(0,2,0)&=6\left(2-\frac{1}{Q_m}-Q_m\right)\,,\nonumber\\
\mathfrak{s}_{3,2}(0,2,1)&=6\left(-Q_m ^3-\frac{1}{Q_m ^3}+12 Q_m ^2+\frac{12}{Q_m ^2}-39 Q_m -\frac{39}{Q_m }+56\right)\,,\nonumber\\
\mathfrak{s}_{3,2}(0,2,2)&=6\left(2 Q_m ^4+\frac{2}{Q_m ^4}-39 Q_m ^3-\frac{39}{Q_m ^3}+208 Q_m ^2+\frac{208}{Q_m ^2}-513
   Q_m -\frac{513}{Q_m }+684\right)\,.
\end{align}}
\subsubsection{Expansion Coefficients for $M=3$ and $N=3$}
For $n=0$ we have for the first few coefficients
{\allowdisplaybreaks\begin{align}
\mathfrak{s}_{3,3}(0,1,0)&=9\left(2-\frac{1}{Q_m}-Q_m\right)\,,\nonumber\\
\mathfrak{s}_{3,3}(0,1,1)&=9\left(2 Q_m ^2+\frac{2}{Q_m ^2}-8 Q_m -\frac{8}{Q_m }+12\right)\,,\nonumber\\
\mathfrak{s}_{3,3}(0,1,2)&=9\left(-Q_m ^3-\frac{1}{Q_m ^3}+12 Q_m ^2+\frac{12}{Q_m ^2}-39 Q_m -\frac{39}{Q_m }+56\right)\,,\nonumber\\
\mathfrak{s}_{3,3}(0,2,0)&=9\left(2-\frac{1}{Q_m}-Q_m\right)\,,\nonumber\\
\mathfrak{s}_{3,3}(0,2,1)&=9\left(-Q_m ^3-\frac{1}{Q_m ^3}+12 Q_m ^2+\frac{12}{Q_m ^2}-39 Q_m -\frac{39}{Q_m }+56\right)\,,\nonumber\\
\mathfrak{s}_{3,3}(0,2,2)&=9\left(2 Q_m ^4+\frac{2}{Q_m ^4}-39 Q_m ^3-\frac{39}{Q_m ^3}+208 Q_m ^2+\frac{208}{Q_m ^2}-513
   Q_m -\frac{513}{Q_m }+684\right)\,.
\end{align}}
For $n=1$ we have explicitly
{\allowdisplaybreaks\begin{align}
\mathfrak{s}_{3,3}(1,1,0)&=\frac{3}{8}\left(Q_m +\frac{1}{Q_m }+4\right)\,,\nonumber\\
\mathfrak{s}_{3,3}(1,1,1)&=\frac{3}{8}\left(4 Q_m ^2+\frac{4}{Q_m ^2}-40 Q_m -\frac{40}{Q_m }+72\right)\,,\nonumber\\
\mathfrak{s}_{3,3}(1,1,2)&=\frac{3}{8}\left(Q_m ^3+\frac{1}{Q_m ^3}+72 Q_m ^2+\frac{72}{Q_m ^2}-369 Q_m -\frac{369}{Q_m }+592\right)\,,\nonumber\\
\mathfrak{s}_{3,3}(1,2,0)&=\frac{3}{8}\left(Q_m +\frac{1}{Q_m }+4\right)\,,\nonumber\\
\mathfrak{s}_{3,3}(1,2,1)&=\frac{3}{8}\left(Q_m ^3+\frac{1}{Q_m ^3}+72 Q_m ^2+\frac{72}{Q_m ^2}-369 Q_m -\frac{369}{Q_m }+592\right)\,,\nonumber\\
\mathfrak{s}_{3,3}(1,2,2)&=\frac{3}{8}\left(4 Q_m ^4+\frac{4}{Q_m ^4}-369 Q_m ^3-\frac{369}{Q_m ^3}+3104 Q_m ^2+\frac{3104}{Q_m ^2}-9327
   Q_m -\frac{9327}{Q_m }+13176\right)\,.
\end{align}}
\subsection{Non-Compact Free Energies}\label{App:NonCompactFreeEnergies}
By direct computation, we find for the first few expansion coefficients $\mathfrak{f}_n^{(\{k_i\})}(\tau,m)$ in equation (\ref{NoncompactExpansionEps}) the following expressions
\begin{align}
&\mathfrak{f}_0^{(1)}=\frac{\varphi_{-2,1}}{4}\,,\hspace{3cm}\mathfrak{f}_1^{(1)}=\frac{1}{2\times 24}\left[\varphi_{0,1}-E_2\varphi_{-2,1}\right]\,,\\
&\mathfrak{f}_2^{(1)}=\frac{1}{10\times 24^2}\left[10\,E_2\varphi_{0,1}-\varphi_{-2,1}\left(5E_2{}^2+13E_4\right)\right]\,,
\end{align}
where $\varphi_{-2,1}(\tau,m)$ and $\varphi_{0,1}(\tau,m)$ are the standard Jacobi forms of weight $-2$ and $0$ respectively and index $1$ and $E_{2k}(\tau)$ (for $k\in\mathbb{N}$) are the weight $2k$ Eisenstein series. For the explicit definitions we refer the reader to \cite{Eichler}. Using a similar notation, we can also tabulate the expressions for $(\{k_i\})=(1,1)$
\begin{align}
\mathfrak{f}^{(1,1)}_0&=\frac{1}{48} \varphi  _{-2,1} \left(E_2 \varphi  _{-2,1}+\varphi  _{0,1}\right)\,,\nonumber\\
\mathfrak{f}^{(1,1)}_1&=\frac{1}{576} \left(\left(E_4-2 E_2{}^2\right) \varphi  _{-2,1}^2+\varphi  _{0,1}^2\right)\,,\nonumber\\
\mathfrak{f}^{(1,1)}_2&=\frac{1}{34560}\left[\left(-10 E_2{}^3-3 E_4  E_2+4 E_6 \right) \varphi  _{-2,1}^2+2 \left(5 E_2{}^2-7 E_4 \right) \varphi  _{0,1} \varphi
   _{-2,1}+5 E_2 \varphi  _{0,1}^2\right]\,,
\end{align}
as well as $(\{k_i\})=(2)$
\begin{align}
&\mathfrak{f}^{(2)}_0=-\frac{1}{24} \left(E_2 -E_2(2\tau)\right) \varphi  _{-2,1}^2\,,\nonumber\\
&\mathfrak{f}^{(2)}_1=\frac{\varphi  _{-2,1} }{2880}\left[\left(15 E_2 {}^2-20 E_2(2\tau){}^2+33 E_4 -28 E_4(2\tau) \right) \varphi  _{-2,1}-20 \left(E_2 -E_2(2\tau) \right)
   \varphi  _{0,1}\right]\,,\nonumber\\
   &\mathfrak{f}^{(2)}_2=\frac{1}{725760}\bigg[\left(280 E_2 {}^3+1659 E_4  E_2 -560 E_2(2\tau) {}^3-2352 E_2(2\tau)  E_4(2\tau) +3149 E_6 -2176 E_6(2\tau) \right) \varphi  _{-2,1}^2\nonumber\\
   &\hspace{1cm}-42
   \left(15 E_2 {}^2-20 E_2(2\tau) {}^2+33 E_4 -28 E_4(2\tau) \right) \varphi  _{0,1} \varphi  _{-2,1}+210 \left(E_2 -E_2(2\tau) \right) \varphi _{0,1}^2\bigg]
\end{align}
For $(\{k_i\})=(1,1,1)$ we have
\begin{align}
&\mathfrak{f}^{(1,1,1)}_0=\frac{\phi _{-2,1} }{576} \left(E_2(1) \phi _{-2,1}+\phi _{0,1}\right){}^2\,,\nonumber\\
&\mathfrak{f}^{(1,1,1)}_1=\frac{1}{6912}\left(E_2(1) \phi _{-2,1}+\phi _{0,1}\right) \left(\left(2 E_4(1)-3 E_2(1){}^2\right) \phi _{-2,1}^2+\phi
   _{0,1}^2\right)\,.
\end{align}
along with For $(\{k_i\})=(2,1)$
\begin{align}
&\mathfrak{f}^{(2,1)}_0=-\frac{1}{384} \left(E_2(1){}^2-E_4(1)\right) \phi _{-2,1}^3\nonumber\\
&\mathfrak{f}^{(2,1)}_1=\frac{\phi _{-2,1}^2 }{4608}\left[\left(3 E_2(1){}^3+5 E_4(1) E_2(1)-8 E_6(1)\right) \phi _{-2,1}-3 \left(E_2(1){}^2-E_4(1)\right) \phi
   _{0,1}\right]
\end{align}
and $(\{k_i\})=(3)$
\begin{align}
&\mathfrak{f}^{(3)}_0=\frac{\phi _{-2,1}^2 }{5760}\left[\left(20 E_2(1){}^2+7 E_4(1)-27 E_4(3)\right) \phi _{-2,1}-30 \left(E_2(1)-E_2(3)\right) \phi
   _{0,1}\right]\,,\nonumber\\
   &\mathfrak{f}^{(3)}_1=-\frac{\phi _{-2,1}}{483840}\bigg[-189 \left(5 E_2(1){}^2-5 E_2(3){}^2+9 E_4(1)-9 E_4(3)\right) \phi _{0,1} \phi _{-2,1}+420
   \left(E_2(1)-E_2(3)\right) \phi _{0,1}^2\nonumber\\
   &+\left(420 E_2(1){}^3+7 \left(281 E_4(1)+513 E_4(3)\right) E_2(1)-12474 E_2(3) E_4(3)+1312 E_6(1)+5184
   E_6(3)\right) \phi _{-2,1}^2\bigg]
\end{align}
where we remind the reader of our notation $E_{2k}=E_{2k}(\tau)$.

 For concreteness, we also list the first few Fourier coefficients in the expansion of $T^{(R)}(\tau,m,\epsilon_1)$, as defined in (\ref{DefExpansionTR}). Indeed, for $R=1$ we have
{\allowdisplaybreaks\begin{align}
\mathfrak{t}_0^{(1)}&=\frac{1}{4}\bigg[2-Q_m -\frac{1}{Q_m }+q
   \left(2 Q_m ^2+\frac{2}{Q_m ^2}-8 Q_m -\frac{8}{Q_m }+12\right)\nonumber\\
   &\hspace{2cm}+q^2 \left(-Q_m ^3-\frac{1}{Q_m ^3}+12 Q_m ^2+\frac{12}{Q_m ^2}-39 Q_m -\frac{39}{Q_m }+56\right)+\ldots\bigg]\,,\nonumber\\
\mathfrak{t}_1^{(1)}&= \frac{1}{24}\bigg[4+Q_m +\frac{1}{Q_m }+q
   \left(4 Q_m ^2+\frac{4}{Q_m ^2}-40 Q_m -\frac{40}{Q_m }+72\right)\nonumber\\
   &\hspace{2cm}+q^2 \left(Q_m ^3+\frac{1}{Q_m ^3}+72 Q_m ^2+\frac{72}{Q_m ^2}-369 Q_m -\frac{369}{Q_m }+592\right)+\ldots\bigg]\,,\nonumber\\
\mathfrak{t}_2^{(1)}&=\frac{1}{1440}\bigg[16+7 Q_m +\frac{7}{Q_m }+q \left(16 Q_m ^2+\frac{16}{Q_m ^2}+536
   Q_m +\frac{536}{Q_m }-1824\right)\nonumber\\
   &\hspace{2cm}+q^2 \left(7 Q_m ^3+\frac{7}{Q_m ^3}-1824 Q_m ^2-\frac{1824}{Q_m ^2}+15873
   Q_m +\frac{15873}{Q_m }-30272\right)+\ldots\bigg]
\end{align}}
while for $R=2$ we obtain
{\allowdisplaybreaks\begin{align}
\mathfrak{t}_0^{(2)}&=\frac{1}{4}\bigg[2-Q_m -\frac{1}{Q_m }+q \left(-Q_m ^3-\frac{1}{Q_m ^3}+12 Q_m ^2+\frac{12}{Q_m ^2}-39
   Q_m -\frac{39}{Q_m }+56\right)\nonumber\\
   &\hspace{1cm}+q^2 \left(2 Q_m ^4+\frac{2}{Q_m ^4}-39 Q_m ^3-\frac{39}{Q_m ^3}+208 Q_m ^2+\frac{208}{Q_m ^2}-513
   Q_m -\frac{513}{Q_m }+684\right)+\ldots\bigg]\,,\nonumber\\
\mathfrak{t}_1^{(2)}&=\frac{1}{24}\bigg[4+Q_m +\frac{1}{Q_m }+q
   \left(Q_m ^3+\frac{1}{Q_m ^3}+72 Q_m ^2+\frac{72}{Q_m ^2}-369
   Q_m -\frac{369}{Q_m }+592\right)\nonumber\\
   &\hspace{1cm}+q^2 \left(4 Q_m ^4+\frac{4}{Q_m ^4}-369 Q_m ^3-\frac{369}{Q_m ^3}+3104
   Q_m ^2+\frac{3104}{Q_m ^2}-9327 Q_m -\frac{9327}{Q_m }+13176\right)+\ldots\bigg]\,,\nonumber\\
\mathfrak{t}_2^{(2)}&=\frac{1}{1440}\bigg[16+7 Q_m +\frac{7}{Q_m }+q \left(7
   Q_m ^3+\frac{7}{Q_m ^3}-1824 Q_m ^2-\frac{1824}{Q_m ^2}+15873
   Q_m +\frac{15873}{Q_m }-30272\right)\nonumber\\
&+q^2 \left(16 Q_m ^4+\frac{16}{Q_m ^4}+15873 Q_m ^3+\frac{15873}{Q_m ^3}-244096
   Q_m ^2-\frac{244096}{Q_m ^2}+925671 Q_m +\frac{925671}{Q_m }-1399968\right)\nonumber\\
   &+\ldots\bigg]\,,
\end{align}}
Finally, the first few terms for the case $R=3$ read
{\allowdisplaybreaks\begin{align}
\mathfrak{t}_0^{(3)}&=\frac{1}{4}\bigg[2-Q_m -\frac{1}{Q_m }+q \left(-8
   Q_m ^3-\frac{8}{Q_m ^3}+56 Q_m ^2+\frac{56}{Q_m ^2}-152
   Q_m -\frac{152}{Q_m }+208\right)\nonumber\\
   &+q^2 \bigg(-Q_m ^5-\frac{1}{Q_m ^5}+56 Q_m ^4+\frac{56}{Q_m ^4}-513 Q_m ^3-\frac{513}{Q_m ^3}+2032
   Q_m ^2+\frac{2032}{Q_m ^2}\nonumber\\
   &\hspace{1cm}-4382 Q_m -\frac{4382}{Q_m }+5616\bigg)+\ldots\bigg]\,,\nonumber\\
\mathfrak{t}_1^{(3)}&=\frac{1}{24}\bigg[4+Q_m +\frac{1}{Q_m }+q \left(-40
   Q_m ^3-\frac{40}{Q_m ^3}+592 Q_m ^2+\frac{592}{Q_m ^2}-2104
   Q_m -\frac{2104}{Q_m }+3104\right)\nonumber\\
   &+q^2 \bigg(Q_m ^5+\frac{1}{Q_m ^5}+592 Q_m ^4+\frac{592}{Q_m ^4}-9327 Q_m ^3-\frac{9327}{Q_m ^3}+47840
   Q_m ^2+\frac{47840}{Q_m ^2}\nonumber\\
   &\hspace{1cm}-117202 Q_m -\frac{117202}{Q_m }+156192\bigg)+\ldots\bigg] \,.
\end{align}}


\end{document}